\documentclass[twocolumn,preprintnumbers,amsmath,amssymb,pra]{revtex4}

\usepackage{graphicx}
\usepackage{dcolumn}
\usepackage{bm}
\usepackage{color}
\usepackage[usenames,dvipsnames]{xcolor}
\usepackage{paralist}
\usepackage{verbatim}
\usepackage[normalem]{ulem}   

\newcommand{\bra}[1]{\langle #1|}
\newcommand{\ket}[1]{|#1\rangle}

\newcommand{\pphi}[0]{\varphi}

\newcommand{\dn}[0]{\downarrow}
\newcommand{\lp}[0]{\left}
\newcommand{\rp}[0]{\right}

\newcommand{\de}[0]{\partial}
\newcommand{\KS}[0]{{K\!S}}

\newcommand{\half}[0]{\frac{1}{2}}
\newcommand{\rarr}[0]{\rightarrow}
\newcommand{\rr}[0]{(\mathbf{r})}

\newcommand{\darsec}[0]{\texttt{DARSEC}}

\newcommand{\footnoteremember}[2]{
\footnote{#2}
\newcounter{#1}
\setcounter{#1}{\value{footnote}}
}
\newcommand{\footnoterecall}[1]{
\footnotemark[\value{#1}]
}

\newcommand{\rb}[0]{\mathbf{r}}

\begin{document}

\title{Elimination of the asymptotic fractional dissociation problem \\ in Kohn-Sham density functional theory using the ensemble-generalization approach}

\author{Eli Kraisler}
\affiliation{Department of Materials and Interfaces, Weizmann Institute of Science, Rehovoth 76100, Israel}

\author{Leeor Kronik}
\affiliation{Department of Materials and Interfaces, Weizmann Institute of Science, Rehovoth 76100, Israel}

\date{\today}

\begin{abstract}

Many approximations within density-functional theory spuriously predict that a many-electron system can dissociate into fractionally charged fragments. Here, we revisit the case of dissociated diatomic molecules, known to exhibit this problem when studied within standard approaches, including the local spin-density approximation (LSDA). By 
employing our recently proposed [E.~Kraisler and L.~Kronik, Phys.\ Rev.\ Lett.~\textbf{110}, 126403 (2013)]
ensemble-generalization we find that asymptotic fractional dissociation is eliminated in all systems examined, even if the underlying exchange-correlation (xc) is still the LSDA. Furthermore, as a result of the ensemble generalization procedure, the Kohn-Sham potential develops a spatial step between the dissociated atoms, reflecting the emergence of the derivative discontinuity in the xc energy functional. This step, predicted in the past for the exact Kohn-Sham potential and observed in some of its more advanced approximate forms, is a desired feature that prevents any fractional charge transfer between the system's fragments. It is usually believed that simple xc approximations such as the LSDA cannot develop this step. Our findings show, however, that ensemble generalization to fractional electron densities automatically introduces the desired step even to the most simple approximate xc functionals and correctly predicts asymptotic integer dissociation.

\end{abstract}

\maketitle

\section{Introduction}

Density-functional theory (DFT)~\cite{HK'64,KS'65,PY,DG,Primer,Perdew09,EngelDreizler2011,Burke12,Capelle_BirdsEye} is a widely-used theoretical framework for studying the electronic properties of matter. It is usually employed by mapping the original $N$ interacting-electron system into an equivalent Kohn-Sham (KS) system of $N$ non-interacting particles subject to a common effective potential, such that the ground-state electron density $n \rr$ is reproduced. While exact in principle, this mapping involves the exchange-correlation (xc) density functional, $E_{xc}[n\rr]$, whose exact form is unknown. Therefore, it is always approximate in practice.

Present-day density-functional approximations (DFAs) already make it widely applicable to a variety of many-electron systems in physics, chemistry, and materials science~\cite{Martin,Hafner,Kaxiras03,Cramer2004,ShollSteckel11,KochHolthausen}.
However, there remain numerous challenges that common DFAs fail to meet. A significant problem, which has both formal and practical implications, is the so-called \emph{problem of fractional dissociation}. It is most easily demonstrated when considering a neutral diatomic molecule, AB. Upon dissociation, the molecule must break up into two neutral atoms, A and B, with an \emph{integer} number of electrons, $N_A^0$ and $N_B^0$, on each~\footnote{If the system possesses special symmetries, fractional occupation can be energetically allowed, as e.g.\ in the case of H$_2^+$, which is discussed below}. 
This observation is known as \emph{principle of integer preference}~\cite{Perdew90}. This fundamental principle is not reproduced by many DFAs. Instead, one finds that a system of two well-separated atoms often reaches its energy minimum when the number of electrons on each of the atoms is fractional: $N_A^0+q$ electrons on atom A and $N_B^0-q$ electrons on atom B, with $q \in (-1,1)$~\cite{PPLB82,Perdew90,Ossowski03,Dutoi06,Gritsenko06,MoriS06,Ruzsinszky06,Vydrov06,Perdew07,Vydrov07}.

The significance of this failure lies beyond the accurate description of dissociation in diatomic systems. It indicates that common DFAs may fail to describe charge transfer in molecules and materials (see, e.g., \cite{Perdew84,Tozer03,Maitra05,Mundt05,Toher05,Koentopp06,Ke07,Hofmann12,Nossa13} and references therein). Furthermore, theoretical analysis of the problem of fractional dissociation allowed an understanding of fundamental properties of many-electron systems~\cite{PPLB82,PerdewLevy83,ShamSchluter83}: If $N$ is allowed to be fractional, the total energy, $E(N)$, of a many-electron system must possess a piecewise-linear dependence on $N$. As a result~\cite{Godby87,Godby88,Chan99,AllenTozer02,Teale08,Harbola98,Mosquera14,Mosquera14a_misc}, $E_{xc}[n \rr]$ may exhibit a discontinuity in its derivative at integer $N$.

It has been shown~\cite{PPLB82,Ruzsinszky06,Ruzsinszky07} that the fractional dissociation problem occurs in approximate DFAs due to the deviation of their energy curves from piecewise linearity: standard (semi-)local calculations for atoms show a convex behavior of the energy $E_a(N_a^0+q)$, where $a \in \{A,B\}$. As a result, the total energy of a well-separated diatomic molecule, which can be expressed as
\begin{equation}\label{eq.EAB}
E_{A...B}(q) = E_A(N_A^0+q) + E_B(N_B^0-q),
\end{equation}
may reach an unphysical minimum at $q \neq 0$ (see Fig.~\ref{fig.E_q_LiF} below). This failure has been discovered as early as 1982 by Perdew {\it et al.}~\cite{PPLB82}, for an infinitely stretched LiH molecule treated within the local density approximation, and has since been found in various molecules with different DFAs~\cite{Ruzsinszky06,Perdew90,Ossowski03,Dutoi06,Gritsenko06,MoriS06,Vydrov06,Perdew07,Vydrov07,Baer10}. Perdew {\it et al.}~\cite{PPLB82} have shown that this would not have happened had the atomic energy curves, and as a result the molecular energy curve, been exactly piecewise linear. Then, in terms of the total energy, the molecular energy curve $E_{A...B}(q)$ would have possessed a non-analytical minimum at $q=0$, obeying the principle of integer preference. In terms of the KS potential, a 'plateau' in the vicinity of one of the atoms would have emerged, shifting the KS potential and the energy levels associated with that atom~\cite{Perdew90,Karolewski09,Tempel09,Makmal11,Fuks11,Hofmann12,Nafziger13,GouldHellgren14}. 
Because all these desired features are absent in the standard local density approximation, spurious transfer of fractional charge is not precluded.

Recently, we have proposed an approximate ensemble generalization of the Hartree-exchange-correlation (Hxc) functional~\cite{Kraisler13}. In contrast to the usual approach, in the generalized approach one does not insert the ensemble density, which integrates to a fractional $N$, into the standard form of the approximate Hxc functional chosen. Instead, the Hxc energy is now evaluated as a weighted sum of Hxc energies obtained using two auxiliary densities, which integrate to integer $N$'s. These are derived from the same self-consistent KS potential by choosing the highest, partially-occupied KS energy level as either fully occupied or completely unoccupied.
We have shown~\cite{Kraisler13,Kraisler14,KraislerSchmidt15} that this ensemble generalization restores, to a large extent, the piecewise-linearity property in the energy and simultaneously introduces an appropriate derivative discontinuity into the xc potential in a natural manner, even when the underlying xc functional is as simple as the local spin-density approximation (LSDA). All this is achieved while neither introducing empiricism nor changing the underlying functional form. This generalization would appear to be of direct relevance to the question of asymptotic fractional dissociation. It is therefore of much interest to examine whether and how it affects this important problem.

Here we employ the ensemble-generalized LSDA, denoted eLSDA, to well-separated diatomic molecules. We find that spurious asymptotic fractional dissociation is eliminated in all systems examined. Furthermore, the KS potential develops the desired spatial step, which explains the absence of spurious fractional charge transfer also from the potential perspective. This shows that, contrary to conventional wisdom, the asymptotic fractional dissociation problem can be eliminated even with simple xc functionals, as long as an appropriate ensemble generalization is used.

\section{Methodology}

In the dissociation limit, a diatomic molecule can be described as two atoms with a possibly fractional electrical charge on each (Eq.~(\ref{eq.EAB}))~\cite{PPLB82}. Generally, a closed system with a fractional number of electrons, $N = N_0 + \alpha$, where $N_0 \in \mathbb{N}$ and $\alpha \in (0,1)$ is not physical, in the sense that electrons do not fracture in chemical systems. A fractional electron number may, however, arise as a time average of an open system, which is free to exchange electrons with its surroundings. Such a system can no longer be described by a pure quantum-mechanical state. Instead, it must be considered as a statistical mixture, or ensemble, of pure (integer electron) states \cite{PPLB82}. Specifically, in our case -- Coulomb systems at zero temperature -- this ensemble state is a linear combination of the pure ground states for $N_0$ and $N_0+1$ electrons, with the statistical weights of $(1-\alpha)$ and $\alpha$, respectively~\cite{DG,Lieb,Cohen12},
~\footnote{Here and below it is assumed that the ground states of the system of interest and of its ion are not degenerate, or that the degeneracy can be lifted by applying an infinitesimal external field},
~\footnote{In the fractional-$N$ systems discussed in this work, the nuclear charges remain unchanged as a function of $\alpha$. This means that the system described by the ensemble, as well at least one of the pure KS-states that comprise this ensemble, integrate to a number of electrons that does not equal the nuclear charge, i.e., represent a charged system. This does not pose any difficulties because the systems are finite. }
.

For describing a quantum system with fractional $N$ in KS-DFT, recall that in KS-DFT one maps the original many-electron system to a fictitious non-interacting one, such that the overall density is retained. In particular, the number of particles in the KS system must equal the number of electrons in the real system. Therefore, the KS density must also integrate to a fractional number of particles and the KS ground state must be an ensemble of $N_0$- and $N_0+1$-states, i.e., $\hat{\Lambda}_\KS = (1-\alpha) \ket{\Phi_{N_0}^{(\alpha)}}\bra{\Phi_{N_0}^{(\alpha)}} + \alpha \ket{\Phi_{N_0+1}^{(\alpha)}}\bra{\Phi_{N_0+1}^{(\alpha)}}$. Both $\ket{\Phi_{N_0}^{(\alpha)}}$ and $\ket{\Phi_{N_0+1}^{(\alpha)}}$ are Slater determinants constructed from one-electron KS orbitals, $\pphi_i^{(\alpha)} \rr$, arising from the \emph{same} KS potential. The difference between the two determinants is that $\ket{\Phi_{N_0}^{(\alpha)}}$ is constructed from $N_0$ orbitals, whereas $\ket{\Phi_{N_0+1}^{(\alpha)}}$ is constructed from $N_0 + 1 $ orbitals.

The above-discussed mapping is enforced by the KS potential, which is the common potential ``felt'' by all KS particles. Therefore, when changing the number of electrons, i.e., changing $\alpha$, the KS potential itself can change because the nature of the mapping is changing. Significant variation of the KS potential with $\alpha$ have been indeed found for both the exact KS potential and various approximations to it (see~\cite{GouldToulouse14} and references therein). As a result, the KS orbitals, $\pphi_i^{(\alpha)}\rr$, and all the quantities derived from them (including $\rho_p^{(\alpha)} \rr$ and $n^{(\alpha)} \rr$ that are introduced below), are generally $\alpha$-dependent. Here we emphasize this by including the superscript $(\alpha)$ explicitly.

The usual approach to the KS-DFT treatment of systems with fractional $N$ is based on the construction of the electron density as
\begin{equation}\label{eq.n}
n^{(\alpha)} \rr = \sum_{i=1}^{N_0} |\pphi_i^{(\alpha)} \rr|^2 + \alpha |\pphi_{N_0+1}^{(\alpha)} \rr|^2,
\end{equation} 
i.e., on occupying the first $N_0$ levels fully but occupying the next one by the electron fraction $\alpha$, so as to obtain $N_0 + \alpha$ electrons overall. 
This density can be equivalently expressed as $n^{(\alpha)} \rr =(1-\alpha) \rho_0^{(\alpha)} \rr+ \alpha \rho_1^{(\alpha)} \rr$, 
where $\rho_p^{(\alpha)} \rr$ is defined as the density due to the first $N_0+p$ KS orbitals: $\rho_p^{(\alpha)}\rr = \sum_{i=1}^{N_0+p} |\pphi_i^{(\alpha)}\rr|^2$, where $p = 0$ or $1$. In other words, the overall density is the weighted sum of the two auxiliary densities obtained from taking the partially occupied orbital as either completely empty or completely full
~\footnote{In our previous publication, Ref.~\cite{Kraisler13}, we defined $\rho_p^{(\alpha)}\rr$ (immediately before Eq.~(3) there) without explicitly denoting its $\alpha$-dependence with the superscript $(\alpha)$. We take this opportunity to correct this oversight.}. These two auxiliary densities are not to be confused with the true densities of the $N_0$ and $N_0+1$ electron systems. In the usual approach, the overall density, $n^{(\alpha)} \rr $, is then employed \emph{directly} in the evaluation of the approximate Hxc energy functional, $E_{Hxc}[n]$, by inserting $n^{(\alpha)} \rr $ into the the functional form used for the description of systems with integer $N$.

Recently, we have shown~\cite{Kraisler13,Kraisler14} that this usual approach, i.e., using the integer Hxc form for systems with fractional $N$, contributes significantly to the spurious deviation from piecewise linearity in the total energy discussed above. To address this problem, we proposed an approximate ensemble-based generalization of the Hxc functional to fractional $N$. For completeness, we provide a short overview of this generalization. Its basic idea is that by considering the ensemble average of the many-electron Coulomb operator, $\hat W = \half \sum_i \sum_{j \neq i} | \rb_i - \rb_j|^{-1}$, in the KS system, the Hxc energy can be generalized as follows:
\begin{equation}\label{eq.ET.gen}
    E_{e-Hxc}[n^{(\alpha)}] = (1-\alpha) E_{Hxc}[\rho_0^{(\alpha)}] + \alpha E_{Hxc}[\rho_1^{(\alpha)}].
\end{equation}
Here, the index $e-$ signifies that the functional is ensemble-generalized, $E_{Hxc}$ is the pure-state Hxc functional mentioned above. The essential difference between the ensemble-generalized functional $E_{e-Hxc}[n^{(\alpha)}]$ and the usual pure-state $E_{Hxc}[n^{(\alpha)}]$ is that we do not linearly combine the pure state densities $\rho_p^{(\alpha)} \rr $ first, and then inserting the result in the Hxc functional. Instead, following the ensemble approach, we first evaluate the Hxc for the pure-state densities separately, then linearly combine the two ensuing $E_{Hxc}[\rho_p^{(\alpha)} \rr]$ energies. 
These two procedures are \emph{not} the same mathematically, because the Hxc functional is not linear with respect to the density.
Only at integer $N$ does the Hxc energy reduce to the same form, namely that of the underlying pure-state Hxc functional.
Because at fractional $N$ the Hxc functional is explicitly dependent on $\rho_p^{(\alpha)} \rr$, i.e., it is explicitly orbital-dependent, treatment within the optimized effective potential (OEP) formalism~\cite{Grabo_MolPhys,EngelDreizler2011,KueKronik08} is required. This is true even if the underlying xc functional is explicitly density-dependent, as is the case with the LSDA.

The generalization in Eq.~(\ref{eq.ET.gen}) is applicable to \emph{any} functional and makes the total energy \emph{explicitly} linear in $\alpha$. However, there may still remain an \emph{implicit} non-linear dependence of the energy on $\alpha$ because of the $\alpha$-dependent KS orbitals. 
In practice, the ensemble-generalization of Eq.(\ref{eq.ET.gen}) has already been shown to significantly improve, but not completely eliminate, deviations from piecewise linearity in the energy of simple atoms and molecules~\cite{Kraisler13}.

Here we focus on infinitely separated diatomic molecules, which can be constructed from the atoms H, Li, C, and F, namely H...H, Li...H, C...H, F...H, Li...Li, Li...F, C...C, C...F and F...F, as well as their ions. All calculations are performed within the LSDA and eLSDA, while varying the number of electrons on each of the constituent atoms, using the \darsec \, code~\cite{Makmal09JCTC}. This code allows for spin-polarized all-electron DFT calculations for single atoms and diatomic molecules, using a real-space prolate-spheroidal grid. In all calculations the total energy and the highest occupied (ho) eigenvalue have been obtained with an accuracy of 0.001 Ry. For all fractional densities, eLSDA calculations were performed within the optimized effective potential (OEP) formalism~\cite{Grabo_MolPhys,EngelDreizler2011,KueKronik08}, usually in the Krieger-Li-Iafrate (KLI) approximation~\cite{KLI}. A special case that required employment of the S-iteration method~\cite{KumPer03a,KumPer03b} for solving the full OEP equation is discussed in the Supplementary Material\footnoteremember{suppl.mat}{See Supplementary Material at http://[to be added by the journal] for details.}.
LSDA results are presented only if the obtained system is bound, which is not always the case for (fractionally) negatively charged ions. For eLSDA, where negative ions were never found to be bound, their energy is taken to be equal to that of a neutral system (plus a fraction of an electron at infinity, whose contribution is zero),~\cite{Tozer07,Teale08} -- as elaborated in the Supplementary Material~\footnoterecall{suppl.mat}.

\section{Results and Discussion}

\begin{figure}[h]
  \includegraphics[trim=0mm 0mm 0mm 0mm,scale=0.67]{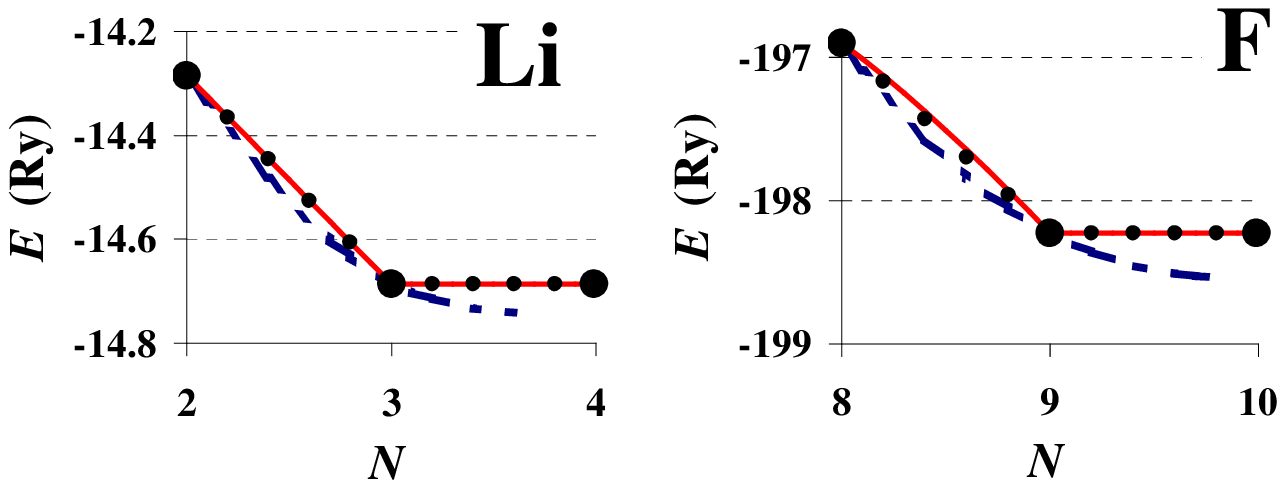}\\
  \vspace{1mm}
  \includegraphics[trim=0mm 0mm 0mm 0mm,scale=0.6]{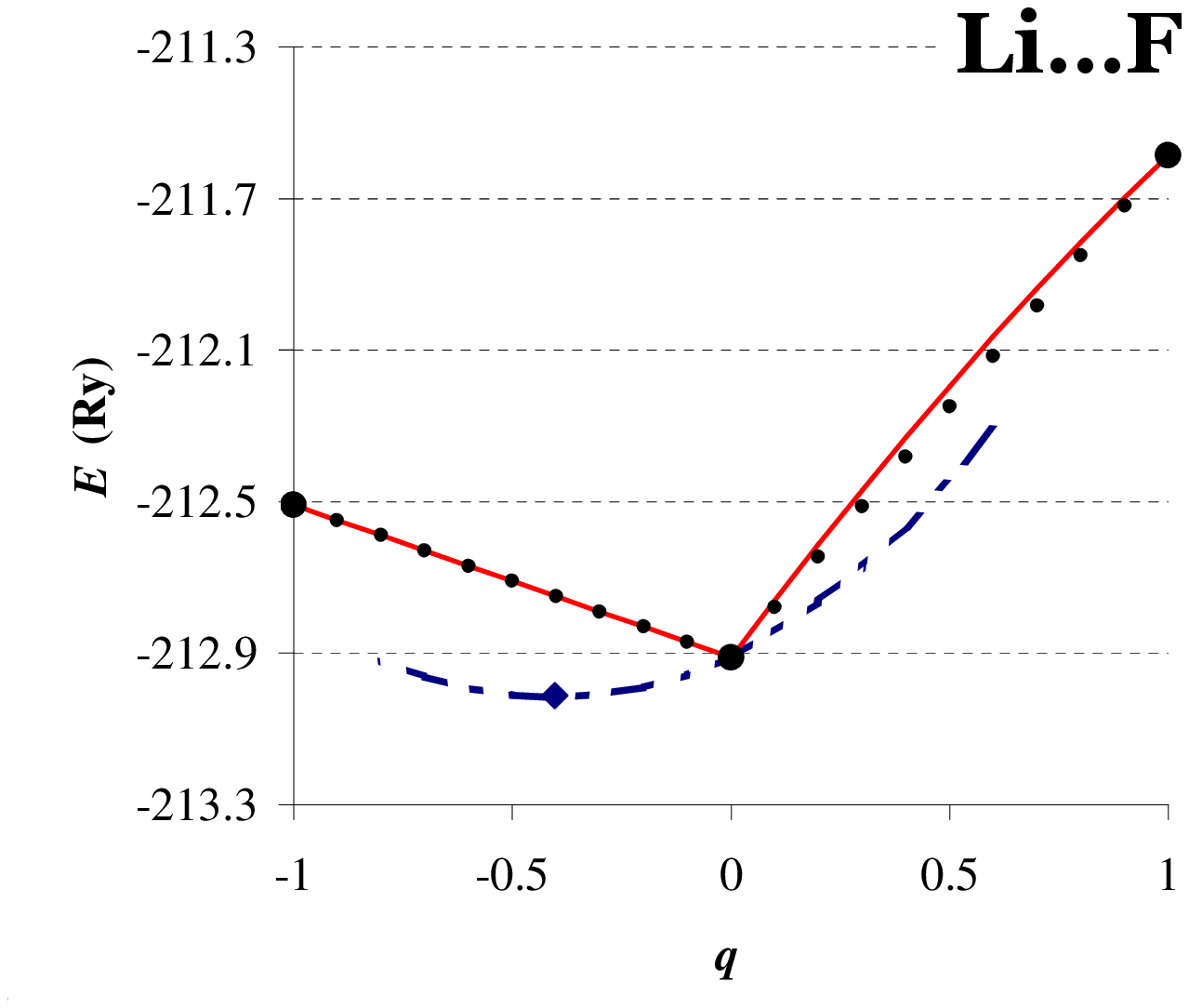}\\
  \caption{(Color online) Total energy of the Li and F atoms as a function of $N$ (top) and energy of the dissociated Li...F molecule as a function of $q$ (bottom), computed with LSDA (dash-dotted line) and eLSDA (solid line), compared to the expected piecewise-linear (dotted line) behavior (the latter is obtained by linear interpolation of the eLSDA energies at integer electron values).}
  \label{fig.E_q_LiF} 
\end{figure}

We first consider the total energy of neutral molecules, $E_{A...B}(q)$, as a function of $q$ -- the amount of negative charge transferred from $B$ to $A$. Figure~\ref{fig.E_q_LiF} presents the total energy of the Li and F atoms as a function of $N$ and the total energy curve of the dissociated molecule Li...F as a function of $q$. The latter curve is obtained from the combination of the atomic curves, according to Eq.~(\ref{eq.EAB}). For the atoms, a clear improvement in restoring piecewise linearity with eLSDA is observed, comparing to LSDA. For the dissociated molecule Li...F, the LSDA produces a convex energy curve, which is far from the expected piecewise-linear behavior, and with a spurious minimum at $q_0 = -0.4$, in agreement with Ref.~\cite{Ossowski03}. In contrast, the eLSDA yields a result that is much closer to the piecewise-linear one, being a combination of two slightly concave curves. It possesses no spurious minimum, but rather a non-analytical minimum at 0, as required. Similar spurious minima are obtained with LSDA for the neutral Li...H, F...H, and C...F molecules, with $q_0=-0.05$, 0.12, and $-0.20$, respectively, all removed by using eLSDA. These additional cases are shown in the Supplementary Material~\footnoterecall{suppl.mat}. Importantly, close inspection of Fig.~\ref{fig.E_q_LiF} (and its analogues in the Supplementary Material) reveals that the energy curves obtained with eLSDA are not perfectly linear but rather are slightly convex. This is a consequence of the remaining implicit $\alpha$-dependence of the e-Hxc functional, owing to the above-discussed $\alpha$-dependence of the KS orbitals, $\pphi_i^{(\alpha)} \rr$. Also for ionized molecules of the type (A...B)$^+$, where $q=0$ corresponds to the state A$^+$...B and $q=1$ to A...B$^+$, with LSDA we find (C...H)$^+$, (F...H)$^+$, and (C...F)$^+$ to possess spurious minima at $q_0=0.43$, 0.67, and 0.27 respectively. With eLSDA all these minima are eliminated. For the homoatomic molecules H...H, Li...Li, C...C and F...F, as well as for C...H, (Li...H)$^+$, (Li...F)$^+$, no spurious minima occur even with the LSDA, but again the eLSDA energy curve is much closer to piecewise-linearity.

Of particular interest are the $E(q)$ curves for homoatomic ionized molecules (A...A)$^+$~\cite{Merkle92,BallySastry97,Ruzsinszky07,Livshits08}.
For these systems we expect the energy to be the sum of the energies of the atom A and the ion A$^+$, independently of $q$. However, here the LSDA incorrectly prefers the state with $q=0.5$, where the electron density is strongly delocalized, i.e., it is present on both nuclei. This is demonstrated, using (H...H)$^+$ and (F...F)$^+$ as examples, in Fig.~\ref{fig.E_q_H2+}, with more examples given in the Supplementary Material \footnoterecall{suppl.mat}. With LSDA, the energy for the $q=0.5$ state is too low by 0.20 and 0.29 Ry, respectively. The eLSDA cures this problem completely for (H...H)$^+$, producing a flat energy curve, as required. 
For (F...F)$^+$, the eLSDA results in a somewhat concave curve, which has its minima at $q=0$ and 1, being too high by 0.1 Ry at $q=0.5$. Thus, integer preference is again enforced. In fact it is ''over-enforced'', in the sense  that the integral states at $q=$0 and 1 are preferred over those with fractional $q$, instead of being iso-energetic with them. This is a consequence of the remaining deviation of the energy curve from a horizontal line, which is again traced to the implicit $\alpha$-dependence of the e-Hxc functional. Still, spurious delocalization is completely eliminated.

\begin{figure}[h]
  \includegraphics[trim=0mm 0mm 0mm 0mm,scale=0.6]{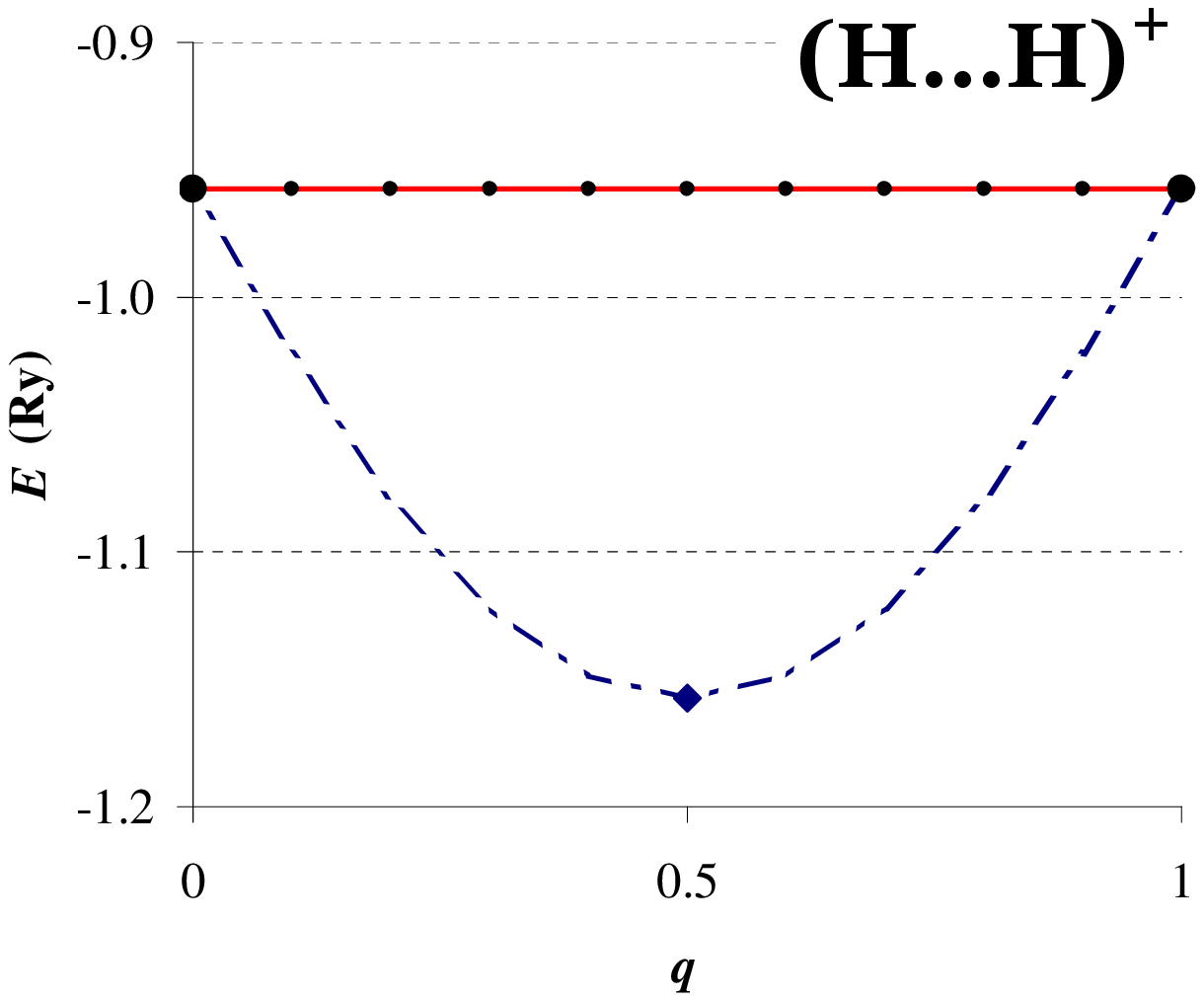}\\
  \vspace{2mm}
  \includegraphics[trim=0mm 0mm 0mm 0mm,scale=0.6]{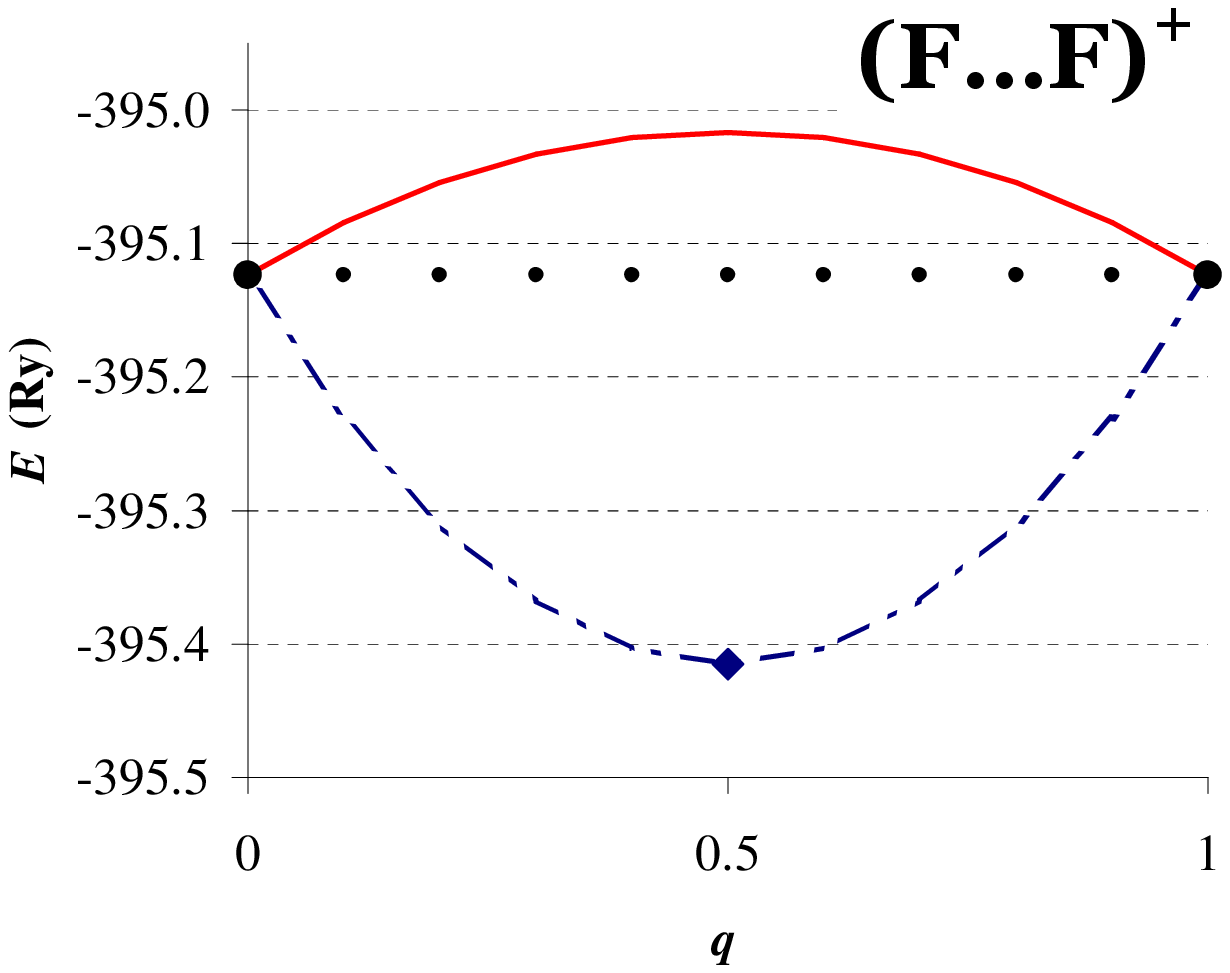}\\
  \caption{(Color online) Total energy of the (H...H)$^+$ and (F...F)$^+$ molecules as a function of $q$ computed with the LSDA (dash-dotted line) and eLSDA (solid line) functionals, compared to the expected piecewise-linear (dotted line) behavior.}
  \label{fig.E_q_H2+} 
\end{figure}

\begin{figure}[h]
  \includegraphics[trim=0mm 0mm 0mm 0mm,scale=0.6]{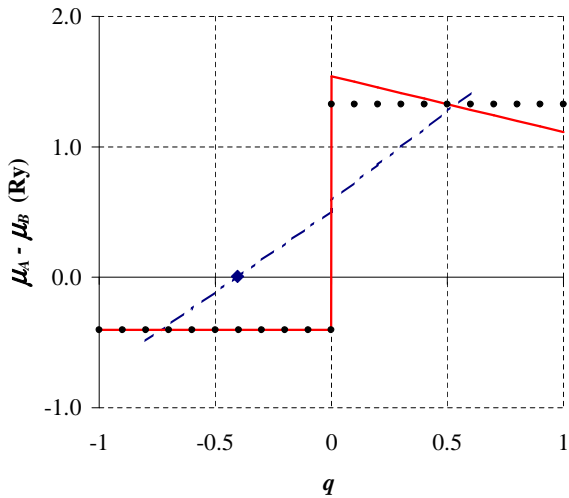}\\
  \caption{(Color online) Dependence of $\mu_A(q) - \mu_B(q)$ on $q$ for the Li...F molecule, computed with LSDA (dash-dotted line) and eLSDA (solid line), compared to the expected discontinuous stair-step (dotted line) behavior.}  
  \label{fig.mu_A_mu_B__LiF} 
\end{figure}

Fractional dissociation and its elimination can be further understood by examining $dE_{A...B}/dq$. From Eq.~(\ref{eq.EAB}) it follows that $dE_{A...B}/dq = \mu_A(q) - \mu_B(q)$, where $\mu_a := dE_a/dN$ is the chemical potential of the $a$-th atom. From the property of piecewise-linearity of the exact $E(N)$ we expect $\mu_A - \mu_B$ to be a stair-step function of $q$. The dependence of $\mu_A - \mu_B$ on $q$ for Li...F is shown in Fig.~\ref{fig.mu_A_mu_B__LiF}.
With LSDA we find that for low enough values of $q$, $\mu_A < \mu_B$. This suggests that charge has to be moved from B to A, thereby increasing $q$. By doing so, we reach a chemical equilibrium for $q_0=-0.4$, with $\mu_A = \mu_B$~\cite{PPLB82}. For higher values of $q$, $\mu_A > \mu_B$, which suggests moving charge back from A to B, returning to the equilibrium at $q_0$. For eLSDA we find a completely different behavior: $\mu_A(q) - \mu_B(q)$ is negative for negative $q$'s, suggesting a transfer of charge from B to A, i.e., increasing $q$, until reaching 0, where a discontinuous change is observed and $\mu_A(q) - \mu_B(q)$ becomes positive, suggesting decreasing $q$ back. This illustrates the fact that $E(q)$ exhibits a non-analytic minimum: while the system reaches its energy minimum at $q=0$, the two atoms are, strictly speaking, never in chemical equilibrium, because $\mu_A \neq \mu_B$. An infinitesimal increase of $q$ above $q=0$ results in an abrupt jump, which indicates that $q$ has to be decreased, and vice versa. Note that the deviation of the eLSDA from the expected stepwise form in Fig.~\ref{fig.mu_A_mu_B__LiF} originates from the residual convexity of the $E(q)$ in Fig.~\ref{fig.E_q_LiF}.

The energy considerations presented above are also reflected in the KS potential. For an infinitely separated diatomic molecule, the molecular potential coincides with the atomic potential in the vicinity of each of the nuclei. Therefore, here it is obtained by combination of the two KS atomic potentials obtained from two separate atomic calculations.

As a result of the ensemble generalization to the energy (Eq.~(\ref{eq.ET.gen})), the KS potential is generalized as well. Taking the functional derivative of $E_{e-Hxc}[n]$ with respect to $n$, while recalling that the ensemble-generalized functional is both $\alpha$- and orbital-dependent~\cite{KueKronik08}, one reveals a peculiar property of the KS potential~\cite{Kraisler13}: it does not necessarily tend asymptotically to 0, but rather to a finite spatial constant, $v_\infty$.
We stress that this constant is well-defined and can be analytically expressed in terms of the KS orbitals and Hxc potential~\cite{Kraisler13,Kraisler14}. It must be taken into account in order for the ho KS eigenvalue to equal $\de E/ \de N$, i.e., to obey Janak's theorem~\cite{Janak}. Indeed, including this constant significantly improves~\cite{Kraisler13,KraislerSchmidt15} the prediction of the ionization potential (IP) from the ho KS eigenvalue via the IP theorem~\cite{PPLB82,Levy84,AlmbladhVonBarth85,PerdewLevy97,Harbola99,Dabo14}. In the current context, we emphasize that $v_\infty$ is system-dependent and may change discontinuously with $N$ across an integer point.

To illustrate how the ensemble generalization prevents spurious charge transfer from the potential perspective, we use the (C...F)$^{++}$ molecule. We prefer it to the Li...F used above to confine ourselves to a system with strictly bound states (see the Supplementary Material~\footnoterecall{suppl.mat} for a complete discussion). 

\begin{figure}
  \includegraphics[trim=20mm 0mm 0mm 0mm,scale=0.30]{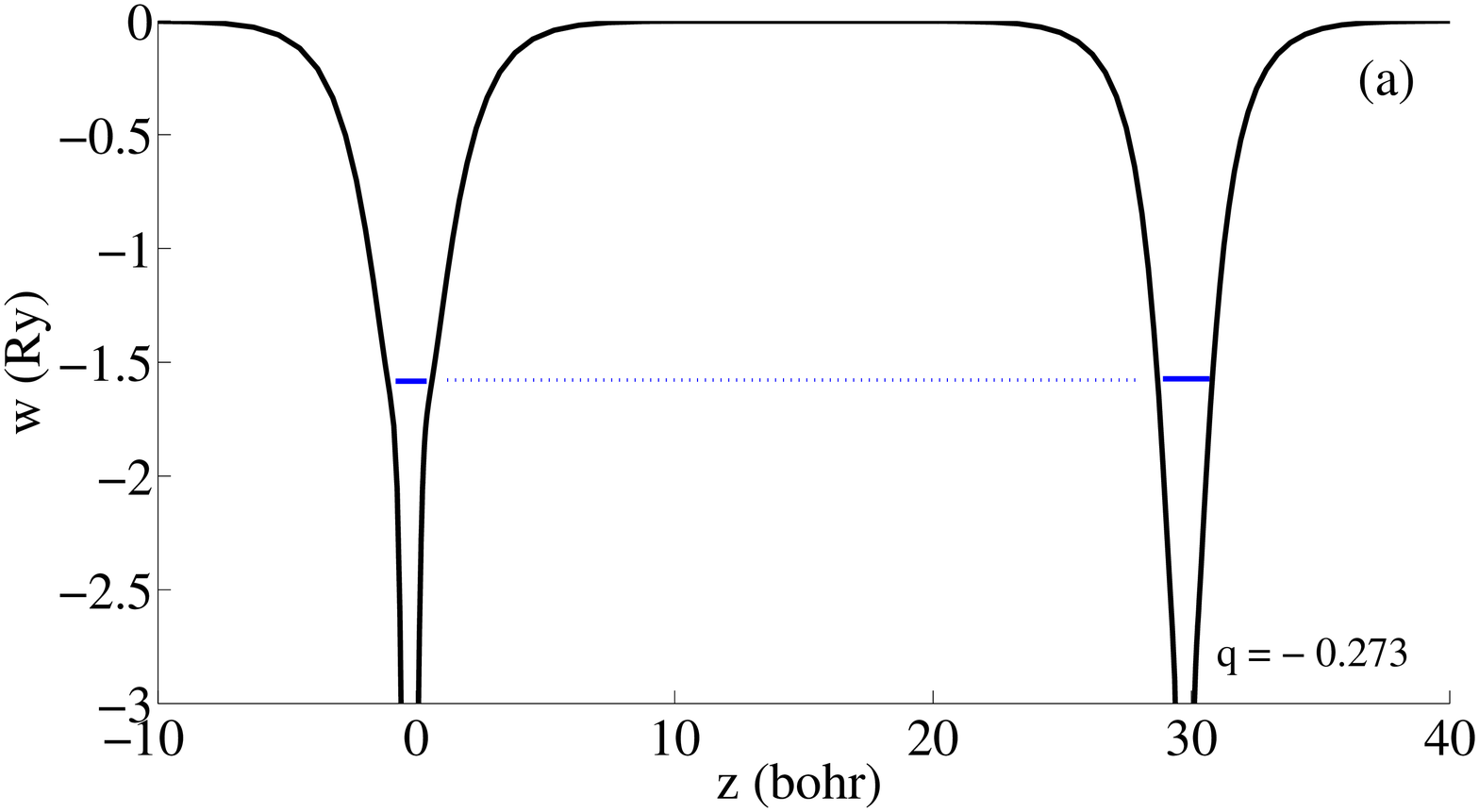}\\
  \includegraphics[trim=20mm 0mm 0mm 0mm,scale=0.30]{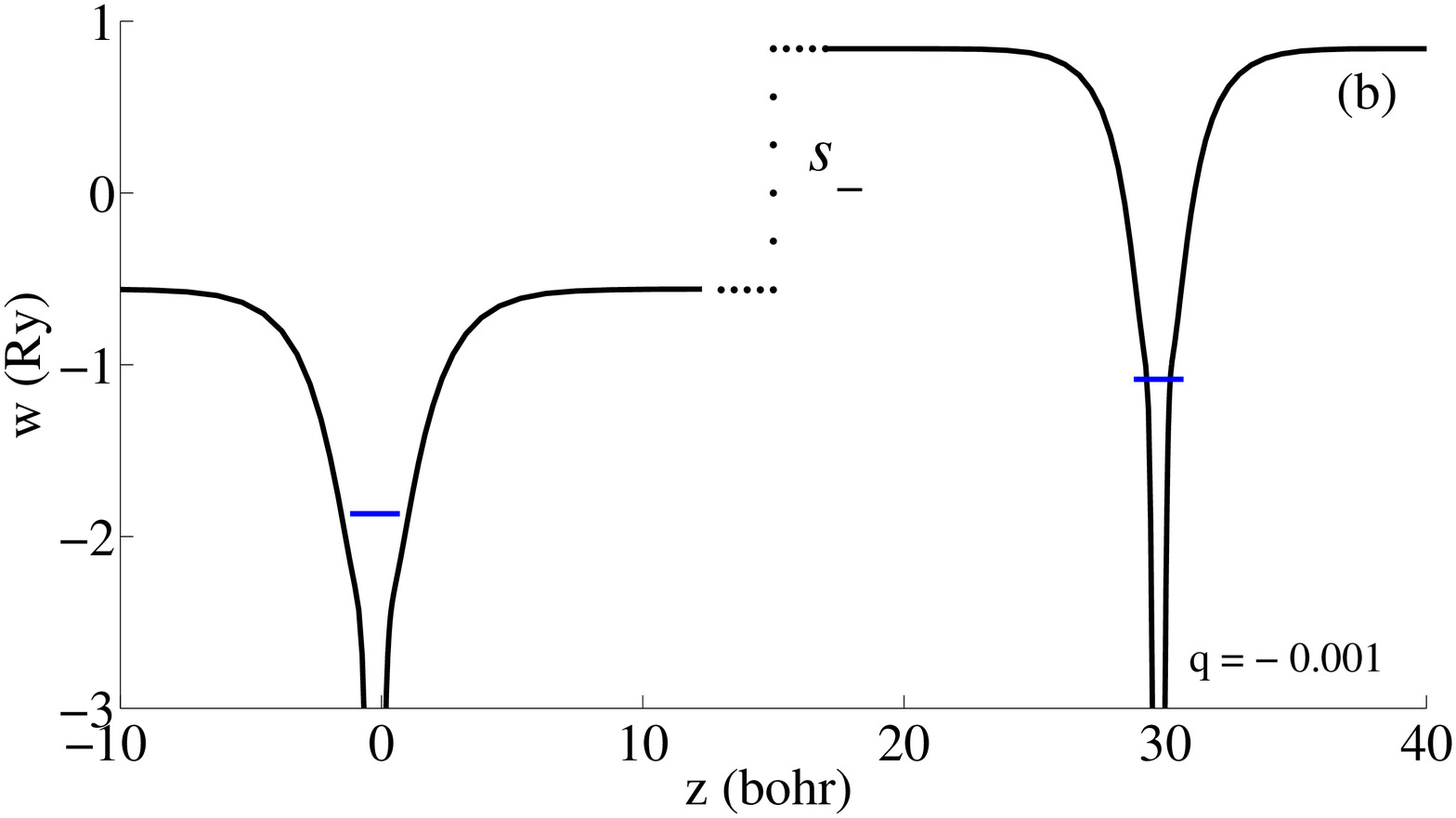}\\
  \includegraphics[trim=20mm 0mm 0mm 0mm,scale=0.30]{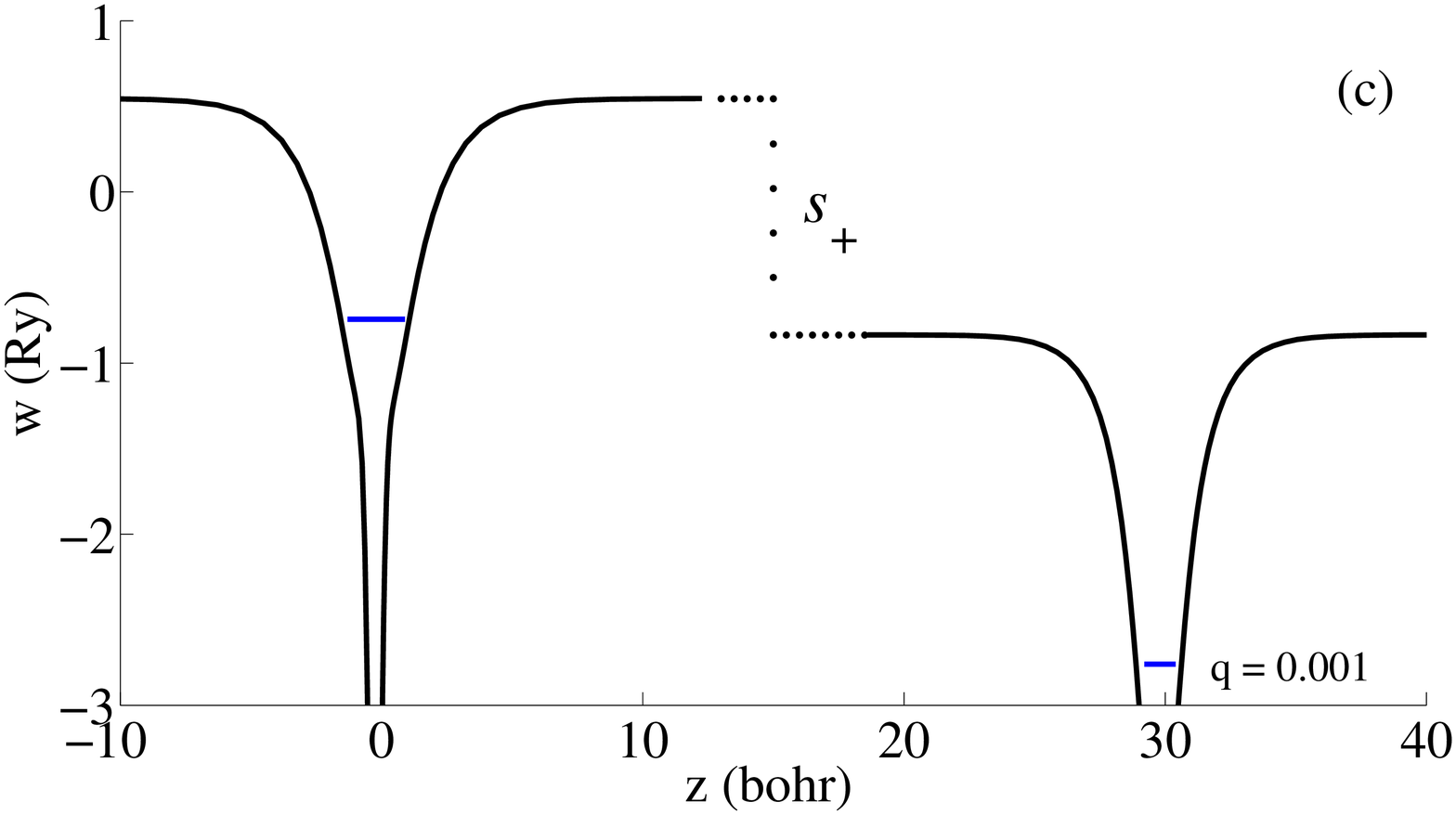}\\
  \caption{(Color online) The potential $w \rr$ for the system C$^{1-q}$...F$^{1+q}$ plotted against the interatomic axis, $z$: (a) with LSDA, at $q=q_0=-0.273$; (b) with eLSDA, at $q=-0.001$; (c) with eLSDA, at $q=+0.001$. The ho levels of both ions are marked with horizontal lines. }
  \label{fig.w} 
\end{figure}

The energy curve of the system C$^{1-q}$...F$^{1+q}$ is qualitatively similar to the one shown at Fig.~\ref{fig.E_q_LiF}, with a spurious minimum at $q_0=-0.273$ for LSDA and a non-analytic minimum at 0 for eLSDA. In the following, we discuss the quantity $w \rr$, which equals the $\dn$-KS potential, minus the external and the Hartree potentials. The latter do not possess any discontinuity and therefore are subtracted for clarity. We stress that Fig.~\ref{fig.w} that depicts $w \rr$ is not at all a schematic drawing. It is obtained by combining  two KS potentials obtained from two separate atomic calculations, for C$^{1-q}$ and F$^{1+q}$, while preserving their asymptotic values
~\footnote{Although we consider a system of two infinitely separated atoms, on the plots they are positioned at a finite distance of 30 Bohr from each other, with the C ion always on the left. Note that the function $w \rr$ is finite everywhere, including near the nuclei. In Fig.~\ref{fig.w} the $w$-axis is cut at $-3$ Ry for clarity of presentation.}.

For LSDA (Fig.~\ref{fig.w}(a)), $w \rr$ essentially differs from 0 only near the two nuclei and decays to 0 elsewhere. For $q<q_0$, the ho level of the C ion is lower than the ho level of the F ion. The levels approach each other gradually as $q$ approaches $q_0$, indicating chemical equilibrium. Further increase of $q$ makes the levels gradually separate. Therefore, chemical equilibrium is achieved by a spurious transfer of a fraction of an electron from the C ion to the F ion, contrary to the principle of integer preference. For eLSDA (Figs.~\ref{fig.w}(b) and~(c)), the situation is radically different. Since the ensemble-generalized KS potential tends asymptotically to a non-zero system-dependent constant, $v_\infty$, the potential experiences a step between the two well-separated ions. Moreover, since $v_\infty$ is discontinuous when $N$ passes through an integer, the value of the step will abruptly change around an integer $q$. For $q<0$, we observe a positive step, $s_-$, so the ho of the C ion is always lower than that of the F ion, suggesting that $q$ should be increased. However, for $q>0$, the step $s_+$ is negative and the ho of the C ion is higher than that of the F ion, suggesting that $q$ should be decreased. This \emph{mechanism of negative reaction} to charge transfer assures that the well-separated subsystems will always be occupied by an integer number of electrons, as expected.
The height of the steps can be expressed analytically in KS quantities of the subsystems A and B with integral $N$: $s_- = v^0_B - v^0_A + \Delta_B$; $s_+ = v^0_B - v^0_A - \Delta_A$, where~\cite{Kraisler14} 
\begin{align}\label{eq.v0.2}
\nonumber     v^0_a = E_{Hxc}[n_a] &- E_{Hxc}[n_a \!\! - \!\! |\pphi_{ho,a}^{(1)}|^2] \\
                    &- \int d^3r |\pphi_{ho,a}^{(1)}\rr|^2 v_{Hxc}[n_a]\rr
\end{align}
is the ensemble-KS-potential asymptote for $N_a \rarr (N_a^0)^-$ (i.e., $\alpha \rarr 1^-$)
~\footnote{Note that the quantity $v^0$ discussed here does \emph{not} necessarily equal the quantity $v_\infty$ mentioned above. They are equal for $N \rarr N_0^-$, but this is not the case for a general value of $N$, including $N \rarr N_0^+$.}, and
\begin{align}\label{eq.Delta}
\nonumber    \Delta_a &= E_{Hxc}[n_a \!\! + \!\! |\pphi_{lu,a}^{(1)}|^2] \!\! - \!\! 2E_{Hxc}[n_a] \!\! + \!\! 
             E_{Hxc}[n_a \!\! - \!\! |\pphi_{ho,a}^{(1)}|^2] \\
           &+ \int  d^3r \,\, \lp( |\pphi_{ho,a}^{(1)}\rr|^2 - |\pphi_{lu,a}^{(1)}\rr|^2 \rp) v_{Hxc}[n_a] \rr
\end{align}
is the ensemble derivative discontinuity of the $a$-th system ($a \in \{A,B\}$) that occurs as $N$ crosses the integer value $N_a^0$. Note that for the exact functional $v^0_a = 0$, i.e., the asymptotic potentials tends to zero, and therefore $s_- = \Delta_B$ and $s_+ = - \Delta_A$.

Importantly, we note that here we only addressed infinitely separated molecules, by treating their constituent atoms independently. Naive application of the above-formulated ensemble generalization to finite bond-length molecules is not useful as the overall number of electrons in the system is still integer, instigating no ensemble correction. Further generalization of the approach, possibly by employing partition-DFT to address sub-systems (see, e.g.,~\cite{Nafziger14} and references therein), so as to overcome this difficulty will be discussed elsewhere.

\section{Conclusions}

In conclusion, from an energy perspective the problem of asymptotic fractional dissociation in DFT emerges owing to significant convexity in the energy versus fractional electron number curves. Here, we have shown that this convexity is replaced by much smaller concavity by employing the ensemble generalization~\cite{Kraisler13}, thereby restoring integer dissociation, even when the underlying xc functional is the simple LSDA. 

From a potential perspective, spurious fractional charge transfer between well-separated molecular fragments occurs in standard approximations because they lack a 'plateau' in the KS potential. This `plateau' is naturally introduced by the ensemble generalization. Therefore, the ensemble generalization enforces the physical integer dissociation limit in both the energy and the potential pictures, without introducing empiricism or altering the underlying functional form.  Finally, we note that integer dissociation is recovered by this procedure despite the fact that both LSDA and eLSDA suffer from one-electron self-interaction. We therefore conclude that the connection between fractional dissociation and one-electron self-interaction~\cite{Perdew90,Zhang98,Pederson14} is not a straightforward one. In our opinion, self-interaction correction~\cite{PZ'81} schemes can indeed remedy fractional dissociation, but this is because they produce a concave energy curve. We expect that the ensemble-generalization perspective can bring significant improvement in the description of charge transfer processes, even with currently existing xc functionals.

\acknowledgements{We thank Prof.\ Kieron Burke (UC Irvine) for encouraging us to address the problem of fractional dissociation. Financial support by the European Research Council and the Lise Meitner center for computational chemistry is gratefully acknowledged. E.K.\ is a recipient of the Levzion scholarship.}

\bibliography{bibliography}

\begin{thebibliography}{78}
\expandafter\ifx\csname natexlab\endcsname\relax\def\natexlab#1{#1}\fi
\expandafter\ifx\csname bibnamefont\endcsname\relax
  \def\bibnamefont#1{#1}\fi
\expandafter\ifx\csname bibfnamefont\endcsname\relax
  \def\bibfnamefont#1{#1}\fi
\expandafter\ifx\csname citenamefont\endcsname\relax
  \def\citenamefont#1{#1}\fi
\expandafter\ifx\csname url\endcsname\relax
  \def\url#1{\texttt{#1}}\fi
\expandafter\ifx\csname urlprefix\endcsname\relax\def\urlprefix{URL }\fi
\providecommand{\bibinfo}[2]{#2}
\providecommand{\eprint}[2][]{\url{#2}}

\bibitem[{\citenamefont{Hohenberg and Kohn}(1964)}]{HK'64}
\bibinfo{author}{\bibfnamefont{P.}~\bibnamefont{Hohenberg}} \bibnamefont{and}
  \bibinfo{author}{\bibfnamefont{W.}~\bibnamefont{Kohn}},
  \bibinfo{journal}{Phys. Rev.} \textbf{\bibinfo{volume}{136}},
  \bibinfo{pages}{B864} (\bibinfo{year}{1964}).

\bibitem[{\citenamefont{Kohn and Sham}(1965)}]{KS'65}
\bibinfo{author}{\bibfnamefont{W.}~\bibnamefont{Kohn}} \bibnamefont{and}
  \bibinfo{author}{\bibfnamefont{L.~J.} \bibnamefont{Sham}},
  \bibinfo{journal}{Phys. Rev.} \textbf{\bibinfo{volume}{140}},
  \bibinfo{pages}{A1133} (\bibinfo{year}{1965}).

\bibitem[{\citenamefont{Parr and Yang}(1989)}]{PY}
\bibinfo{author}{\bibfnamefont{R.~G.} \bibnamefont{Parr}} \bibnamefont{and}
  \bibinfo{author}{\bibfnamefont{W.}~\bibnamefont{Yang}},
  \emph{\bibinfo{title}{Density-Functional Theory of Atoms and Molecules}}
  (\bibinfo{publisher}{Oxford University Press}, \bibinfo{year}{1989}).

\bibitem[{\citenamefont{{R.M. Dreizler} and {E.K.U. Gross}}(1990)}]{DG}
\bibinfo{author}{\bibnamefont{{R.M. Dreizler}}} \bibnamefont{and}
  \bibinfo{author}{\bibnamefont{{E.K.U. Gross}}},
  \emph{\bibinfo{title}{{D}ensity {F}unctional {T}heory}}
  (\bibinfo{publisher}{Springer Verlag}, \bibinfo{address}{Berlin},
  \bibinfo{year}{1990}).

\bibitem[{\citenamefont{Fiolhais et~al.}(2003)\citenamefont{Fiolhais, Nogueira,
  and Marques}}]{Primer}
\bibinfo{editor}{\bibfnamefont{C.}~\bibnamefont{Fiolhais}},
  \bibinfo{editor}{\bibfnamefont{F.}~\bibnamefont{Nogueira}}, \bibnamefont{and}
  \bibinfo{editor}{\bibfnamefont{M.~A.} \bibnamefont{Marques}}, eds.,
  \emph{\bibinfo{title}{A Primer in Density Functional Theory}}
  (\bibinfo{publisher}{Springer}, \bibinfo{year}{2003}), vol.
  \bibinfo{volume}{620} of \emph{\bibinfo{series}{Lectures in Physics}}.

\bibitem[{\citenamefont{Perdew et~al.}(2009)\citenamefont{Perdew, Ruzsinszky,
  Constantin, Sun, and Csonka}}]{Perdew09}
\bibinfo{author}{\bibfnamefont{J.~P.} \bibnamefont{Perdew}},
  \bibinfo{author}{\bibfnamefont{A.}~\bibnamefont{Ruzsinszky}},
  \bibinfo{author}{\bibfnamefont{L.~A.} \bibnamefont{Constantin}},
  \bibinfo{author}{\bibfnamefont{J.}~\bibnamefont{Sun}}, \bibnamefont{and}
  \bibinfo{author}{\bibfnamefont{G.~I.} \bibnamefont{Csonka}},
  \bibinfo{journal}{J. Chem. Theory Comp.} \textbf{\bibinfo{volume}{5}},
  \bibinfo{pages}{902} (\bibinfo{year}{2009}).

\bibitem[{\citenamefont{Engel and Dreizler}(2011)}]{EngelDreizler2011}
\bibinfo{author}{\bibfnamefont{E.}~\bibnamefont{Engel}} \bibnamefont{and}
  \bibinfo{author}{\bibfnamefont{R.}~\bibnamefont{Dreizler}},
  \emph{\bibinfo{title}{Density Functional Theory: An Advanced Course}}
  (\bibinfo{publisher}{Springer}, \bibinfo{year}{2011}).

\bibitem[{\citenamefont{Burke}(2012)}]{Burke12}
\bibinfo{author}{\bibfnamefont{K.}~\bibnamefont{Burke}}, \bibinfo{journal}{J.
  Chem. Phys.} \textbf{\bibinfo{volume}{136}}, \bibinfo{pages}{150901}
  (\bibinfo{year}{2012}).

\bibitem[{\citenamefont{Capelle}(2006)}]{Capelle_BirdsEye}
\bibinfo{author}{\bibfnamefont{K.}~\bibnamefont{Capelle}},
  \bibinfo{journal}{Braz. J. Phys.} \textbf{\bibinfo{volume}{36}},
  \bibinfo{pages}{1318} (\bibinfo{year}{2006}).

\bibitem[{\citenamefont{Martin}(2004)}]{Martin}
\bibinfo{author}{\bibfnamefont{R.}~\bibnamefont{Martin}},
  \emph{\bibinfo{title}{Electronic Structure}} (\bibinfo{publisher}{Cambridge
  Unviersity Press}, \bibinfo{year}{2004}).

\bibitem[{\citenamefont{Hafner}(2000)}]{Hafner}
\bibinfo{author}{\bibfnamefont{J.}~\bibnamefont{Hafner}},
  \bibinfo{journal}{Acta Mater.} \textbf{\bibinfo{volume}{48}},
  \bibinfo{pages}{71 } (\bibinfo{year}{2000}).

\bibitem[{\citenamefont{Kaxiras}(2003)}]{Kaxiras03}
\bibinfo{author}{\bibfnamefont{E.}~\bibnamefont{Kaxiras}},
  \emph{\bibinfo{title}{{Atomic and Electronic Structure of Solids}}}
  (\bibinfo{publisher}{Cambridge University Press}, \bibinfo{year}{2003}).

\bibitem[{\citenamefont{Cramer}(2004)}]{Cramer2004}
\bibinfo{author}{\bibfnamefont{C.}~\bibnamefont{Cramer}},
  \emph{\bibinfo{title}{Essentials Of Computational Chemistry: Theories And
  Models}} (\bibinfo{publisher}{Wiley}, \bibinfo{year}{2004}).

\bibitem[{\citenamefont{Sholl and Steckel}(2011)}]{ShollSteckel11}
\bibinfo{author}{\bibfnamefont{D.}~\bibnamefont{Sholl}} \bibnamefont{and}
  \bibinfo{author}{\bibfnamefont{J.}~\bibnamefont{Steckel}},
  \emph{\bibinfo{title}{Density Functional Theory: A Practical Introduction}}
  (\bibinfo{publisher}{Wiley}, \bibinfo{year}{2011}).

\bibitem[{\citenamefont{Koch and Holthausen}(2000)}]{KochHolthausen}
\bibinfo{author}{\bibfnamefont{W.}~\bibnamefont{Koch}} \bibnamefont{and}
  \bibinfo{author}{\bibfnamefont{M.}~\bibnamefont{Holthausen}},
  \emph{\bibinfo{title}{A chemist's guide to density functional theory}}
  (\bibinfo{publisher}{Wiley-VCH}, \bibinfo{year}{2000}).

\bibitem[{\citenamefont{Perdew}(1990)}]{Perdew90}
\bibinfo{author}{\bibfnamefont{J.~P.} \bibnamefont{Perdew}},
  \bibinfo{journal}{Adv. Quantum Chem.} \textbf{\bibinfo{volume}{21}},
  \bibinfo{pages}{113} (\bibinfo{year}{1990}).

\bibitem[{\citenamefont{Perdew et~al.}(1982)\citenamefont{Perdew, Parr, Levy,
  and Balduz}}]{PPLB82}
\bibinfo{author}{\bibfnamefont{J.~P.} \bibnamefont{Perdew}},
  \bibinfo{author}{\bibfnamefont{R.~G.} \bibnamefont{Parr}},
  \bibinfo{author}{\bibfnamefont{M.}~\bibnamefont{Levy}}, \bibnamefont{and}
  \bibinfo{author}{\bibfnamefont{J.~L.} \bibnamefont{Balduz}},
  \bibinfo{journal}{Phys. Rev. Lett.} \textbf{\bibinfo{volume}{49}},
  \bibinfo{pages}{1691} (\bibinfo{year}{1982}).

\bibitem[{\citenamefont{Ossowski et~al.}(2003)\citenamefont{Ossowski, Boyer,
  Mehl, and Pederson}}]{Ossowski03}
\bibinfo{author}{\bibfnamefont{M.}~\bibnamefont{Ossowski}},
  \bibinfo{author}{\bibfnamefont{L.}~\bibnamefont{Boyer}},
  \bibinfo{author}{\bibfnamefont{M.}~\bibnamefont{Mehl}}, \bibnamefont{and}
  \bibinfo{author}{\bibfnamefont{M.}~\bibnamefont{Pederson}},
  \bibinfo{journal}{Phys. Rev. B} \textbf{\bibinfo{volume}{68}},
  \bibinfo{pages}{245107} (\bibinfo{year}{2003}).

\bibitem[{\citenamefont{Dutoi and Head-Gordon}(2006)}]{Dutoi06}
\bibinfo{author}{\bibfnamefont{A.~D.} \bibnamefont{Dutoi}} \bibnamefont{and}
  \bibinfo{author}{\bibfnamefont{M.}~\bibnamefont{Head-Gordon}},
  \bibinfo{journal}{Chem. Phys. Lett.} \textbf{\bibinfo{volume}{422}},
  \bibinfo{pages}{230} (\bibinfo{year}{2006}).

\bibitem[{\citenamefont{Gritsenko and Baerends}(2006)}]{Gritsenko06}
\bibinfo{author}{\bibfnamefont{O.}~\bibnamefont{Gritsenko}} \bibnamefont{and}
  \bibinfo{author}{\bibfnamefont{E.}~\bibnamefont{Baerends}},
  \bibinfo{journal}{Int. J. Quantum Chem.} \textbf{\bibinfo{volume}{106}},
  \bibinfo{pages}{3167} (\bibinfo{year}{2006}).

\bibitem[{\citenamefont{Mori-S\'{a}nchez
  et~al.}(2006)\citenamefont{Mori-S\'{a}nchez, Cohen, and Yang}}]{MoriS06}
\bibinfo{author}{\bibfnamefont{P.}~\bibnamefont{Mori-S\'{a}nchez}},
  \bibinfo{author}{\bibfnamefont{A.~J.} \bibnamefont{Cohen}}, \bibnamefont{and}
  \bibinfo{author}{\bibfnamefont{W.}~\bibnamefont{Yang}}, \bibinfo{journal}{J.
  Chem. Phys.} \textbf{\bibinfo{volume}{125}}, \bibinfo{pages}{201102}
  (\bibinfo{year}{2006}).

\bibitem[{\citenamefont{Ruzsinszky et~al.}(2006)\citenamefont{Ruzsinszky,
  Perdew, Csonka, Vydrov, and Scuseria}}]{Ruzsinszky06}
\bibinfo{author}{\bibfnamefont{A.}~\bibnamefont{Ruzsinszky}},
  \bibinfo{author}{\bibfnamefont{J.~P.} \bibnamefont{Perdew}},
  \bibinfo{author}{\bibfnamefont{G.~I.} \bibnamefont{Csonka}},
  \bibinfo{author}{\bibfnamefont{O.~A.} \bibnamefont{Vydrov}},
  \bibnamefont{and} \bibinfo{author}{\bibfnamefont{G.~E.}
  \bibnamefont{Scuseria}}, \bibinfo{journal}{J. Chem. Phys.}
  \textbf{\bibinfo{volume}{125}}, \bibinfo{pages}{194112}
  (\bibinfo{year}{2006}).

\bibitem[{\citenamefont{Vydrov and Scuseria}(2006)}]{Vydrov06}
\bibinfo{author}{\bibfnamefont{O.~A.} \bibnamefont{Vydrov}} \bibnamefont{and}
  \bibinfo{author}{\bibfnamefont{G.~E.} \bibnamefont{Scuseria}},
  \bibinfo{journal}{J. Chem. Phys.} \textbf{\bibinfo{volume}{125}},
  \bibinfo{pages}{234109} (\bibinfo{year}{2006}).

\bibitem[{\citenamefont{Perdew et~al.}(2007)\citenamefont{Perdew, Ruzsinszky,
  Csonka, Vydrov, Scuseria, Staroverov, and Tao}}]{Perdew07}
\bibinfo{author}{\bibfnamefont{J.}~\bibnamefont{Perdew}},
  \bibinfo{author}{\bibfnamefont{A.}~\bibnamefont{Ruzsinszky}},
  \bibinfo{author}{\bibfnamefont{G.}~\bibnamefont{Csonka}},
  \bibinfo{author}{\bibfnamefont{O.}~\bibnamefont{Vydrov}},
  \bibinfo{author}{\bibfnamefont{G.}~\bibnamefont{Scuseria}},
  \bibinfo{author}{\bibfnamefont{V.}~\bibnamefont{Staroverov}},
  \bibnamefont{and} \bibinfo{author}{\bibfnamefont{J.}~\bibnamefont{Tao}},
  \bibinfo{journal}{Phys. Rev. A} \textbf{\bibinfo{volume}{76}},
  \bibinfo{pages}{040501} (\bibinfo{year}{2007}).

\bibitem[{\citenamefont{Vydrov et~al.}(2007)\citenamefont{Vydrov, Scuseria, and
  Perdew}}]{Vydrov07}
\bibinfo{author}{\bibfnamefont{O.~A.} \bibnamefont{Vydrov}},
  \bibinfo{author}{\bibfnamefont{G.~E.} \bibnamefont{Scuseria}},
  \bibnamefont{and} \bibinfo{author}{\bibfnamefont{J.~P.}
  \bibnamefont{Perdew}}, \bibinfo{journal}{J. Chem. Phys.}
  \textbf{\bibinfo{volume}{126}}, \bibinfo{pages}{154109}
  (\bibinfo{year}{2007}).

\bibitem[{\citenamefont{Perdew and Smith}(1984)}]{Perdew84}
\bibinfo{author}{\bibfnamefont{J.~P.} \bibnamefont{Perdew}} \bibnamefont{and}
  \bibinfo{author}{\bibfnamefont{J.}~\bibnamefont{Smith}},
  \bibinfo{journal}{Surf. Sci.} \textbf{\bibinfo{volume}{141}},
  \bibinfo{pages}{L295} (\bibinfo{year}{1984}).

\bibitem[{\citenamefont{Tozer}(2003)}]{Tozer03}
\bibinfo{author}{\bibfnamefont{D.~J.} \bibnamefont{Tozer}},
  \bibinfo{journal}{J. Chem. Phys.} \textbf{\bibinfo{volume}{119}},
  \bibinfo{pages}{12697} (\bibinfo{year}{2003}).

\bibitem[{\citenamefont{Maitra}(2005)}]{Maitra05}
\bibinfo{author}{\bibfnamefont{N.~T.} \bibnamefont{Maitra}},
  \bibinfo{journal}{J. Chem. Phys.} \textbf{\bibinfo{volume}{122}},
  \bibinfo{pages}{234104} (\bibinfo{year}{2005}).

\bibitem[{\citenamefont{Mundt and K\"{u}mmel}(2005)}]{Mundt05}
\bibinfo{author}{\bibfnamefont{M.}~\bibnamefont{Mundt}} \bibnamefont{and}
  \bibinfo{author}{\bibfnamefont{S.}~\bibnamefont{K\"{u}mmel}},
  \bibinfo{journal}{Phys. Rev. Lett.} \textbf{\bibinfo{volume}{95}},
  \bibinfo{pages}{203004} (\bibinfo{year}{2005}).

\bibitem[{\citenamefont{Toher et~al.}(2005)\citenamefont{Toher, Filippetti,
  Sanvito, and Burke}}]{Toher05}
\bibinfo{author}{\bibfnamefont{C.}~\bibnamefont{Toher}},
  \bibinfo{author}{\bibfnamefont{A.}~\bibnamefont{Filippetti}},
  \bibinfo{author}{\bibfnamefont{S.}~\bibnamefont{Sanvito}}, \bibnamefont{and}
  \bibinfo{author}{\bibfnamefont{K.}~\bibnamefont{Burke}},
  \bibinfo{journal}{Phys. Rev. Lett.} \textbf{\bibinfo{volume}{95}},
  \bibinfo{pages}{146402} (\bibinfo{year}{2005}).

\bibitem[{\citenamefont{Koentopp et~al.}(2006)\citenamefont{Koentopp, Burke,
  and Evers}}]{Koentopp06}
\bibinfo{author}{\bibfnamefont{M.}~\bibnamefont{Koentopp}},
  \bibinfo{author}{\bibfnamefont{K.}~\bibnamefont{Burke}}, \bibnamefont{and}
  \bibinfo{author}{\bibfnamefont{F.}~\bibnamefont{Evers}},
  \bibinfo{journal}{Phys. Rev. B} \textbf{\bibinfo{volume}{73}},
  \bibinfo{pages}{121403} (\bibinfo{year}{2006}).

\bibitem[{\citenamefont{Ke et~al.}(2007)\citenamefont{Ke, Baranger, and
  Yang}}]{Ke07}
\bibinfo{author}{\bibfnamefont{S.-H.} \bibnamefont{Ke}},
  \bibinfo{author}{\bibfnamefont{H.~U.} \bibnamefont{Baranger}},
  \bibnamefont{and} \bibinfo{author}{\bibfnamefont{W.}~\bibnamefont{Yang}},
  \bibinfo{journal}{J. Chem. Phys.} \textbf{\bibinfo{volume}{126}},
  \bibinfo{pages}{201102} (\bibinfo{year}{2007}).

\bibitem[{\citenamefont{Hofmann and K\"{u}mmel}(2012)}]{Hofmann12}
\bibinfo{author}{\bibfnamefont{D.}~\bibnamefont{Hofmann}} \bibnamefont{and}
  \bibinfo{author}{\bibfnamefont{S.}~\bibnamefont{K\"{u}mmel}},
  \bibinfo{journal}{Phys. Rev. B} \textbf{\bibinfo{volume}{86}},
  \bibinfo{pages}{201109} (\bibinfo{year}{2012}).

\bibitem[{\citenamefont{Nossa et~al.}(2013)\citenamefont{Nossa, Islam, Canali,
  and Pederson}}]{Nossa13}
\bibinfo{author}{\bibfnamefont{J.~F.} \bibnamefont{Nossa}},
  \bibinfo{author}{\bibfnamefont{M.~F.} \bibnamefont{Islam}},
  \bibinfo{author}{\bibfnamefont{C.~M.} \bibnamefont{Canali}},
  \bibnamefont{and} \bibinfo{author}{\bibfnamefont{M.~R.}
  \bibnamefont{Pederson}}, \bibinfo{journal}{Phys. Rev. B}
  \textbf{\bibinfo{volume}{88}}, \bibinfo{pages}{224423}
  (\bibinfo{year}{2013}).

\bibitem[{\citenamefont{Perdew and Levy}(1983)}]{PerdewLevy83}
\bibinfo{author}{\bibfnamefont{J.~P.} \bibnamefont{Perdew}} \bibnamefont{and}
  \bibinfo{author}{\bibfnamefont{M.}~\bibnamefont{Levy}},
  \bibinfo{journal}{Phys. Rev. Lett.} \textbf{\bibinfo{volume}{51}},
  \bibinfo{pages}{1884} (\bibinfo{year}{1983}).

\bibitem[{\citenamefont{Sham and Schl\"{u}ter}(1983)}]{ShamSchluter83}
\bibinfo{author}{\bibfnamefont{L.}~\bibnamefont{Sham}} \bibnamefont{and}
  \bibinfo{author}{\bibfnamefont{M.}~\bibnamefont{Schl\"{u}ter}},
  \bibinfo{journal}{Phys. Rev. Lett.} \textbf{\bibinfo{volume}{51}},
  \bibinfo{pages}{1888} (\bibinfo{year}{1983}).

\bibitem[{\citenamefont{Godby et~al.}(1987)\citenamefont{Godby, Schl\"{u}ter,
  and Sham}}]{Godby87}
\bibinfo{author}{\bibfnamefont{R.}~\bibnamefont{Godby}},
  \bibinfo{author}{\bibfnamefont{M.}~\bibnamefont{Schl\"{u}ter}},
  \bibnamefont{and} \bibinfo{author}{\bibfnamefont{L.}~\bibnamefont{Sham}},
  \bibinfo{journal}{Phys. Rev. B} \textbf{\bibinfo{volume}{36}},
  \bibinfo{pages}{6497} (\bibinfo{year}{1987}).

\bibitem[{\citenamefont{Godby et~al.}(1988)\citenamefont{Godby, Schl\"{u}ter,
  and Sham}}]{Godby88}
\bibinfo{author}{\bibfnamefont{R.}~\bibnamefont{Godby}},
  \bibinfo{author}{\bibfnamefont{M.}~\bibnamefont{Schl\"{u}ter}},
  \bibnamefont{and} \bibinfo{author}{\bibfnamefont{L.}~\bibnamefont{Sham}},
  \bibinfo{journal}{Phys. Rev. B} \textbf{\bibinfo{volume}{37}},
  \bibinfo{pages}{10159} (\bibinfo{year}{1988}).

\bibitem[{\citenamefont{Chan}(1999)}]{Chan99}
\bibinfo{author}{\bibfnamefont{G.~K.-L.} \bibnamefont{Chan}},
  \bibinfo{journal}{J. Chem. Phys.} \textbf{\bibinfo{volume}{110}},
  \bibinfo{pages}{4710} (\bibinfo{year}{1999}).

\bibitem[{\citenamefont{Allen and Tozer}(2002)}]{AllenTozer02}
\bibinfo{author}{\bibfnamefont{M.}~\bibnamefont{Allen}} \bibnamefont{and}
  \bibinfo{author}{\bibfnamefont{D.}~\bibnamefont{Tozer}},
  \bibinfo{journal}{Mol.Phys.} \textbf{\bibinfo{volume}{100}},
  \bibinfo{pages}{433} (\bibinfo{year}{2002}).

\bibitem[{\citenamefont{Teale et~al.}(2008)\citenamefont{Teale, de~Proft, and
  Tozer}}]{Teale08}
\bibinfo{author}{\bibfnamefont{A.~M.} \bibnamefont{Teale}},
  \bibinfo{author}{\bibfnamefont{F.}~\bibnamefont{de~Proft}}, \bibnamefont{and}
  \bibinfo{author}{\bibfnamefont{D.~J.} \bibnamefont{Tozer}},
  \bibinfo{journal}{J. Chem. Phys.} \textbf{\bibinfo{volume}{129}},
  \bibinfo{pages}{044110} (\bibinfo{year}{2008}).

\bibitem[{\citenamefont{Harbola}(1998)}]{Harbola98}
\bibinfo{author}{\bibfnamefont{M.~K.} \bibnamefont{Harbola}},
  \bibinfo{journal}{Phys. Rev. A} \textbf{\bibinfo{volume}{57}},
  \bibinfo{pages}{4253} (\bibinfo{year}{1998}).

\bibitem[{\citenamefont{Mosquera and Wasserman}(2014)}]{Mosquera14}
\bibinfo{author}{\bibfnamefont{M.~A.} \bibnamefont{Mosquera}} \bibnamefont{and}
  \bibinfo{author}{\bibfnamefont{A.}~\bibnamefont{Wasserman}},
  \bibinfo{journal}{Phys. Rev. A} \textbf{\bibinfo{volume}{89}},
  \bibinfo{pages}{052506} (\bibinfo{year}{2014}).

\bibitem[{\citenamefont{Mosquera and Wasserman}()}]{Mosquera14a_misc}
\bibinfo{author}{\bibfnamefont{M.~A.} \bibnamefont{Mosquera}} \bibnamefont{and}
  \bibinfo{author}{\bibfnamefont{A.}~\bibnamefont{Wasserman}},
  \bibinfo{note}{{M}ol. Phys. (2014); doi: 10.1080/00268976.2014.968650}.

\bibitem[{\citenamefont{Ruzsinszky et~al.}(2007)\citenamefont{Ruzsinszky,
  Perdew, Csonka, Vydrov, and Scuseria}}]{Ruzsinszky07}
\bibinfo{author}{\bibfnamefont{A.}~\bibnamefont{Ruzsinszky}},
  \bibinfo{author}{\bibfnamefont{J.~P.} \bibnamefont{Perdew}},
  \bibinfo{author}{\bibfnamefont{G.~I.} \bibnamefont{Csonka}},
  \bibinfo{author}{\bibfnamefont{O.~A.} \bibnamefont{Vydrov}},
  \bibnamefont{and} \bibinfo{author}{\bibfnamefont{G.~E.}
  \bibnamefont{Scuseria}}, \bibinfo{journal}{J. Chem. Phys.}
  \textbf{\bibinfo{volume}{126}}, \bibinfo{pages}{104102}
  (\bibinfo{year}{2007}).

\bibitem[{\citenamefont{Baer et~al.}(2010)\citenamefont{Baer, Livshits, and
  Salzner}}]{Baer10}
\bibinfo{author}{\bibfnamefont{R.}~\bibnamefont{Baer}},
  \bibinfo{author}{\bibfnamefont{E.}~\bibnamefont{Livshits}}, \bibnamefont{and}
  \bibinfo{author}{\bibfnamefont{U.}~\bibnamefont{Salzner}},
  \bibinfo{journal}{Annu. Rev. Phys. Chem.} \textbf{\bibinfo{volume}{61}},
  \bibinfo{pages}{85} (\bibinfo{year}{2010}).

\bibitem[{\citenamefont{Karolewski et~al.}(2009)\citenamefont{Karolewski,
  Armiento, and K\"{u}mmel}}]{Karolewski09}
\bibinfo{author}{\bibfnamefont{A.}~\bibnamefont{Karolewski}},
  \bibinfo{author}{\bibfnamefont{R.}~\bibnamefont{Armiento}}, \bibnamefont{and}
  \bibinfo{author}{\bibfnamefont{S.}~\bibnamefont{K\"{u}mmel}},
  \bibinfo{journal}{J. Chem. Theory Comp.} \textbf{\bibinfo{volume}{5}},
  \bibinfo{pages}{712} (\bibinfo{year}{2009}).

\bibitem[{\citenamefont{Tempel et~al.}(2009)\citenamefont{Tempel, Martinez, and
  Maitra}}]{Tempel09}
\bibinfo{author}{\bibfnamefont{D.}~\bibnamefont{Tempel}},
  \bibinfo{author}{\bibfnamefont{T.}~\bibnamefont{Martinez}}, \bibnamefont{and}
  \bibinfo{author}{\bibfnamefont{N.}~\bibnamefont{Maitra}},
  \bibinfo{journal}{J. Chem. Theory and Comput.} \textbf{\bibinfo{volume}{5}},
  \bibinfo{pages}{770} (\bibinfo{year}{2009}).

\bibitem[{\citenamefont{Makmal et~al.}(2011)\citenamefont{Makmal, K\"{u}mmel,
  and Kronik}}]{Makmal11}
\bibinfo{author}{\bibfnamefont{A.}~\bibnamefont{Makmal}},
  \bibinfo{author}{\bibfnamefont{S.}~\bibnamefont{K\"{u}mmel}},
  \bibnamefont{and} \bibinfo{author}{\bibfnamefont{L.}~\bibnamefont{Kronik}},
  \bibinfo{journal}{Phys. Rev. A} \textbf{\bibinfo{volume}{83}},
  \bibinfo{pages}{062512} (\bibinfo{year}{2011}).

\bibitem[{\citenamefont{Fuks et~al.}(2011)\citenamefont{Fuks, Rubio, and
  Maitra}}]{Fuks11}
\bibinfo{author}{\bibfnamefont{J.~I.} \bibnamefont{Fuks}},
  \bibinfo{author}{\bibfnamefont{A.}~\bibnamefont{Rubio}}, \bibnamefont{and}
  \bibinfo{author}{\bibfnamefont{N.~T.} \bibnamefont{Maitra}},
  \bibinfo{journal}{Phys. Rev. A} \textbf{\bibinfo{volume}{83}},
  \bibinfo{pages}{042501} (\bibinfo{year}{2011}).

\bibitem[{\citenamefont{Nafziger and Wasserman}(2013)}]{Nafziger13}
\bibinfo{author}{\bibfnamefont{J.}~\bibnamefont{Nafziger}} \bibnamefont{and}
  \bibinfo{author}{\bibfnamefont{A.}~\bibnamefont{Wasserman}}
  (\bibinfo{year}{2013}), \eprint{arXiv:1305.4966v2}.

\bibitem[{\citenamefont{Gould and Hellgren}(2014)}]{GouldHellgren14}
\bibinfo{author}{\bibfnamefont{T.}~\bibnamefont{Gould}} \bibnamefont{and}
  \bibinfo{author}{\bibfnamefont{M.}~\bibnamefont{Hellgren}}
  (\bibinfo{year}{2014}), \eprint{arXiv:1406.7609v1}.

\bibitem[{\citenamefont{Kraisler and Kronik}(2013)}]{Kraisler13}
\bibinfo{author}{\bibfnamefont{E.}~\bibnamefont{Kraisler}} \bibnamefont{and}
  \bibinfo{author}{\bibfnamefont{L.}~\bibnamefont{Kronik}},
  \bibinfo{journal}{Phys. Rev. Lett.} \textbf{\bibinfo{volume}{110}},
  \bibinfo{pages}{126403} (\bibinfo{year}{2013}).

\bibitem[{\citenamefont{Kraisler and Kronik}(2014)}]{Kraisler14}
\bibinfo{author}{\bibfnamefont{E.}~\bibnamefont{Kraisler}} \bibnamefont{and}
  \bibinfo{author}{\bibfnamefont{L.}~\bibnamefont{Kronik}},
  \bibinfo{journal}{J. Chem. Phys.} \textbf{\bibinfo{volume}{140}},
  \bibinfo{pages}{18A540} (\bibinfo{year}{2014}).

\bibitem[{\citenamefont{Kraisler et~al.}()\citenamefont{Kraisler, Schmidt,
  K\"{u}mmel, and Kronik}}]{KraislerSchmidt15}
\bibinfo{author}{\bibfnamefont{E.}~\bibnamefont{Kraisler}},
  \bibinfo{author}{\bibfnamefont{T.}~\bibnamefont{Schmidt}},
  \bibinfo{author}{\bibfnamefont{S.}~\bibnamefont{K\"{u}mmel}},
  \bibnamefont{and} \bibinfo{author}{\bibfnamefont{L.}~\bibnamefont{Kronik}},
  \bibinfo{note}{in preparation}.

\bibitem[{\citenamefont{Lieb}(1983)}]{Lieb}
\bibinfo{author}{\bibfnamefont{E.~H.} \bibnamefont{Lieb}},
  \bibinfo{journal}{Int. J. Quantum Chem.} \textbf{\bibinfo{volume}{24}},
  \bibinfo{pages}{243} (\bibinfo{year}{1983}).

\bibitem[{\citenamefont{Cohen et~al.}(2012)\citenamefont{Cohen,
  Mori-S\'{a}nchez, and Yang}}]{Cohen12}
\bibinfo{author}{\bibfnamefont{A.~J.} \bibnamefont{Cohen}},
  \bibinfo{author}{\bibfnamefont{P.}~\bibnamefont{Mori-S\'{a}nchez}},
  \bibnamefont{and} \bibinfo{author}{\bibfnamefont{W.}~\bibnamefont{Yang}},
  \bibinfo{journal}{Chem. Rev.} \textbf{\bibinfo{volume}{112}},
  \bibinfo{pages}{289} (\bibinfo{year}{2012}).

\bibitem[{\citenamefont{Gould and Toulouse}(2014)}]{GouldToulouse14}
\bibinfo{author}{\bibfnamefont{T.}~\bibnamefont{Gould}} \bibnamefont{and}
  \bibinfo{author}{\bibfnamefont{J.}~\bibnamefont{Toulouse}},
  \bibinfo{journal}{Phys. Rev. A} \textbf{\bibinfo{volume}{90}},
  \bibinfo{pages}{050502(R)} (\bibinfo{year}{2014}).

\bibitem[{\citenamefont{Grabo et~al.}(1997)\citenamefont{Grabo, Kreibich, and
  Gross}}]{Grabo_MolPhys}
\bibinfo{author}{\bibfnamefont{T.}~\bibnamefont{Grabo}},
  \bibinfo{author}{\bibfnamefont{T.}~\bibnamefont{Kreibich}}, \bibnamefont{and}
  \bibinfo{author}{\bibfnamefont{E.~K.~U.} \bibnamefont{Gross}},
  \bibinfo{journal}{Mol. Eng.} \textbf{\bibinfo{volume}{7}},
  \bibinfo{pages}{27} (\bibinfo{year}{1997}).

\bibitem[{\citenamefont{K\"ummel and Kronik}(2008)}]{KueKronik08}
\bibinfo{author}{\bibfnamefont{S.}~\bibnamefont{K\"ummel}} \bibnamefont{and}
  \bibinfo{author}{\bibfnamefont{L.}~\bibnamefont{Kronik}},
  \bibinfo{journal}{Rev. Mod. Phys.} \textbf{\bibinfo{volume}{80}},
  \bibinfo{pages}{3} (\bibinfo{year}{2008}).

\bibitem[{\citenamefont{Makmal et~al.}(2009)\citenamefont{Makmal, K\"ummel, and
  Kronik}}]{Makmal09JCTC}
\bibinfo{author}{\bibfnamefont{A.}~\bibnamefont{Makmal}},
  \bibinfo{author}{\bibfnamefont{S.}~\bibnamefont{K\"ummel}}, \bibnamefont{and}
  \bibinfo{author}{\bibfnamefont{L.}~\bibnamefont{Kronik}},
  \bibinfo{journal}{J. Chem. Theory Comput.} \textbf{\bibinfo{volume}{5}},
  \bibinfo{pages}{1731} (\bibinfo{year}{2009}).

\bibitem[{\citenamefont{Li et~al.}(1993)\citenamefont{Li, Krieger, and
  Iafrate}}]{KLI}
\bibinfo{author}{\bibfnamefont{Y.}~\bibnamefont{Li}},
  \bibinfo{author}{\bibfnamefont{J.~B.} \bibnamefont{Krieger}},
  \bibnamefont{and} \bibinfo{author}{\bibfnamefont{G.~J.}
  \bibnamefont{Iafrate}}, \bibinfo{journal}{Phys. Rev. A}
  \textbf{\bibinfo{volume}{47}}, \bibinfo{pages}{165} (\bibinfo{year}{1993}).

\bibitem[{\citenamefont{K\"{u}mmel and Perdew}(2003)}]{KumPer03a}
\bibinfo{author}{\bibfnamefont{S.}~\bibnamefont{K\"{u}mmel}} \bibnamefont{and}
  \bibinfo{author}{\bibfnamefont{J.}~\bibnamefont{Perdew}},
  \bibinfo{journal}{Phys. Rev. Lett.} \textbf{\bibinfo{volume}{90}},
  \bibinfo{pages}{043004} (\bibinfo{year}{2003}).

\bibitem[{\citenamefont{K\"ummel and Perdew}(2003)}]{KumPer03b}
\bibinfo{author}{\bibfnamefont{S.}~\bibnamefont{K\"ummel}} \bibnamefont{and}
  \bibinfo{author}{\bibfnamefont{J.~P.} \bibnamefont{Perdew}},
  \bibinfo{journal}{Phys. Rev. B} \textbf{\bibinfo{volume}{68}},
  \bibinfo{pages}{035103} (\bibinfo{year}{2003}).

\bibitem[{\citenamefont{Tozer and {De Proft}}(2007)}]{Tozer07}
\bibinfo{author}{\bibfnamefont{D.~J.} \bibnamefont{Tozer}} \bibnamefont{and}
  \bibinfo{author}{\bibfnamefont{F.}~\bibnamefont{{De Proft}}},
  \bibinfo{journal}{J. Chem. Phys.} \textbf{\bibinfo{volume}{127}},
  \bibinfo{pages}{034108} (\bibinfo{year}{2007}).

\bibitem[{\citenamefont{Merkle et~al.}(1992)\citenamefont{Merkle, Savin, and
  Preuss}}]{Merkle92}
\bibinfo{author}{\bibfnamefont{R.}~\bibnamefont{Merkle}},
  \bibinfo{author}{\bibfnamefont{A.}~\bibnamefont{Savin}}, \bibnamefont{and}
  \bibinfo{author}{\bibfnamefont{H.}~\bibnamefont{Preuss}},
  \bibinfo{journal}{J. Chem. Phys.} \textbf{\bibinfo{volume}{97}},
  \bibinfo{pages}{9216} (\bibinfo{year}{1992}).

\bibitem[{\citenamefont{Bally and Sastry}(1997)}]{BallySastry97}
\bibinfo{author}{\bibfnamefont{T.}~\bibnamefont{Bally}} \bibnamefont{and}
  \bibinfo{author}{\bibfnamefont{G.~N.} \bibnamefont{Sastry}},
  \bibinfo{journal}{J. Phys. Chem. A} \textbf{\bibinfo{volume}{101}},
  \bibinfo{pages}{7923} (\bibinfo{year}{1997}).

\bibitem[{\citenamefont{Livshits and Baer}(2008)}]{Livshits08}
\bibinfo{author}{\bibfnamefont{E.}~\bibnamefont{Livshits}} \bibnamefont{and}
  \bibinfo{author}{\bibfnamefont{R.}~\bibnamefont{Baer}}, \bibinfo{journal}{J.
  Phys. Chem. A} \textbf{\bibinfo{volume}{112}}, \bibinfo{pages}{12789}
  (\bibinfo{year}{2008}).

\bibitem[{\citenamefont{Janak}(1978)}]{Janak}
\bibinfo{author}{\bibfnamefont{J.~F.} \bibnamefont{Janak}},
  \bibinfo{journal}{Phys. Rev. B} \textbf{\bibinfo{volume}{18}},
  \bibinfo{pages}{7165} (\bibinfo{year}{1978}).

\bibitem[{\citenamefont{Levy et~al.}(1984)\citenamefont{Levy, Perdew, and
  Sahni}}]{Levy84}
\bibinfo{author}{\bibfnamefont{M.}~\bibnamefont{Levy}},
  \bibinfo{author}{\bibfnamefont{J.~P.} \bibnamefont{Perdew}},
  \bibnamefont{and} \bibinfo{author}{\bibfnamefont{V.}~\bibnamefont{Sahni}},
  \bibinfo{journal}{Phys. Rev. A} \textbf{\bibinfo{volume}{30}},
  \bibinfo{pages}{2745} (\bibinfo{year}{1984}).

\bibitem[{\citenamefont{Almbladh and {Von Barth}}(1985)}]{AlmbladhVonBarth85}
\bibinfo{author}{\bibfnamefont{C.}~\bibnamefont{Almbladh}} \bibnamefont{and}
  \bibinfo{author}{\bibfnamefont{U.}~\bibnamefont{{Von Barth}}},
  \bibinfo{journal}{Phys. Rev. B} \textbf{\bibinfo{volume}{31}},
  \bibinfo{pages}{3231} (\bibinfo{year}{1985}).

\bibitem[{\citenamefont{Perdew and Levy}(1997)}]{PerdewLevy97}
\bibinfo{author}{\bibfnamefont{J.~P.} \bibnamefont{Perdew}} \bibnamefont{and}
  \bibinfo{author}{\bibfnamefont{M.}~\bibnamefont{Levy}},
  \bibinfo{journal}{Phys. Rev. B} \textbf{\bibinfo{volume}{56}},
  \bibinfo{pages}{16021} (\bibinfo{year}{1997}).

\bibitem[{\citenamefont{Harbola}(1999)}]{Harbola99}
\bibinfo{author}{\bibfnamefont{M.~K.} \bibnamefont{Harbola}},
  \bibinfo{journal}{Phys. Rev. B} \textbf{\bibinfo{volume}{60}},
  \bibinfo{pages}{4545} (\bibinfo{year}{1999}).

\bibitem[{\citenamefont{Dabo et~al.}(2014)\citenamefont{Dabo, Ferretti, and
  Marzari}}]{Dabo14}
\bibinfo{author}{\bibfnamefont{I.}~\bibnamefont{Dabo}},
  \bibinfo{author}{\bibfnamefont{A.}~\bibnamefont{Ferretti}}, \bibnamefont{and}
  \bibinfo{author}{\bibfnamefont{N.}~\bibnamefont{Marzari}},
  \bibinfo{journal}{Top. Curr. Chem.} \textbf{\bibinfo{volume}{347}},
  \bibinfo{pages}{193} (\bibinfo{year}{2014}).

\bibitem[{\citenamefont{Nafziger and Wasserman}(2014)}]{Nafziger14}
\bibinfo{author}{\bibfnamefont{J.}~\bibnamefont{Nafziger}} \bibnamefont{and}
  \bibinfo{author}{\bibfnamefont{A.}~\bibnamefont{Wasserman}},
  \bibinfo{journal}{J. Phys. Chem. A} \textbf{\bibinfo{volume}{118}},
  \bibinfo{pages}{7623} (\bibinfo{year}{2014}).

\bibitem[{\citenamefont{Zhang and Yang}(1998)}]{Zhang98}
\bibinfo{author}{\bibfnamefont{Y.}~\bibnamefont{Zhang}} \bibnamefont{and}
  \bibinfo{author}{\bibfnamefont{W.}~\bibnamefont{Yang}},
  \textbf{\bibinfo{volume}{109}}, \bibinfo{pages}{2604} (\bibinfo{year}{1998}).

\bibitem[{\citenamefont{Pederson et~al.}(2014)\citenamefont{Pederson,
  Ruzsinszky, and Perdew}}]{Pederson14}
\bibinfo{author}{\bibfnamefont{M.~R.} \bibnamefont{Pederson}},
  \bibinfo{author}{\bibfnamefont{A.}~\bibnamefont{Ruzsinszky}},
  \bibnamefont{and} \bibinfo{author}{\bibfnamefont{J.~P.}
  \bibnamefont{Perdew}}, \bibinfo{journal}{J. Chem. Phys.}
  \textbf{\bibinfo{volume}{140}}, \bibinfo{pages}{121103}
  (\bibinfo{year}{2014}).

\bibitem[{\citenamefont{Perdew and Zunger}(1981)}]{PZ'81}
\bibinfo{author}{\bibfnamefont{J.~P.} \bibnamefont{Perdew}} \bibnamefont{and}
  \bibinfo{author}{\bibfnamefont{A.}~\bibnamefont{Zunger}},
  \bibinfo{journal}{Phys. Rev. B} \textbf{\bibinfo{volume}{23}},
  \bibinfo{pages}{5048} (\bibinfo{year}{1981}).

\end{thebibliography}


\begin{thebibliography}{9}
\expandafter\ifx\csname natexlab\endcsname\relax\def\natexlab#1{#1}\fi
\expandafter\ifx\csname bibnamefont\endcsname\relax
  \def\bibnamefont#1{#1}\fi
\expandafter\ifx\csname bibfnamefont\endcsname\relax
  \def\bibfnamefont#1{#1}\fi
\expandafter\ifx\csname citenamefont\endcsname\relax
  \def\citenamefont#1{#1}\fi
\expandafter\ifx\csname url\endcsname\relax
  \def\url#1{\texttt{#1}}\fi
\expandafter\ifx\csname urlprefix\endcsname\relax\def\urlprefix{URL }\fi
\providecommand{\bibinfo}[2]{#2}
\providecommand{\eprint}[2][]{\url{#2}}

\bibitem[{\citenamefont{Tozer and {De Proft}}(2007)}]{Tozer07}
\bibinfo{author}{\bibfnamefont{D.~J.} \bibnamefont{Tozer}} \bibnamefont{and}
  \bibinfo{author}{\bibfnamefont{F.}~\bibnamefont{{De Proft}}},
  \bibinfo{journal}{J. Chem. Phys.} \textbf{\bibinfo{volume}{127}},
  \bibinfo{pages}{034108} (\bibinfo{year}{2007}).

\bibitem[{\citenamefont{Teale et~al.}(2008)\citenamefont{Teale, de~Proft, and
  Tozer}}]{Teale08}
\bibinfo{author}{\bibfnamefont{A.~M.} \bibnamefont{Teale}},
  \bibinfo{author}{\bibfnamefont{F.}~\bibnamefont{de~Proft}}, \bibnamefont{and}
  \bibinfo{author}{\bibfnamefont{D.~J.} \bibnamefont{Tozer}},
  \bibinfo{journal}{J. Chem. Phys.} \textbf{\bibinfo{volume}{129}},
  \bibinfo{pages}{044110} (\bibinfo{year}{2008}).

\bibitem[{\citenamefont{Fiolhais et~al.}(2003)\citenamefont{Fiolhais, Nogueira,
  and Marques}}]{Primer}
\bibinfo{editor}{\bibfnamefont{C.}~\bibnamefont{Fiolhais}},
  \bibinfo{editor}{\bibfnamefont{F.}~\bibnamefont{Nogueira}}, \bibnamefont{and}
  \bibinfo{editor}{\bibfnamefont{M.~A.} \bibnamefont{Marques}}, eds.,
  \emph{\bibinfo{title}{A Primer in Density Functional Theory}}
  (\bibinfo{publisher}{Springer}, \bibinfo{year}{2003}), vol.
  \bibinfo{volume}{620} of \emph{\bibinfo{series}{Lectures in Physics}}.

\bibitem[{\citenamefont{Mori-S\'{a}nchez
  et~al.}(2006)\citenamefont{Mori-S\'{a}nchez, Cohen, and Yang}}]{MoriS06}
\bibinfo{author}{\bibfnamefont{P.}~\bibnamefont{Mori-S\'{a}nchez}},
  \bibinfo{author}{\bibfnamefont{A.~J.} \bibnamefont{Cohen}}, \bibnamefont{and}
  \bibinfo{author}{\bibfnamefont{W.}~\bibnamefont{Yang}}, \bibinfo{journal}{J.
  Chem. Phys.} \textbf{\bibinfo{volume}{125}}, \bibinfo{pages}{201102}
  (\bibinfo{year}{2006}).

\bibitem[{\citenamefont{Vydrov et~al.}(2007)\citenamefont{Vydrov, Scuseria, and
  Perdew}}]{Vydrov07}
\bibinfo{author}{\bibfnamefont{O.~A.} \bibnamefont{Vydrov}},
  \bibinfo{author}{\bibfnamefont{G.~E.} \bibnamefont{Scuseria}},
  \bibnamefont{and} \bibinfo{author}{\bibfnamefont{J.~P.}
  \bibnamefont{Perdew}}, \bibinfo{journal}{J. Chem. Phys.}
  \textbf{\bibinfo{volume}{126}}, \bibinfo{pages}{154109}
  (\bibinfo{year}{2007}).

\bibitem[{\citenamefont{Cohen et~al.}(2012)\citenamefont{Cohen,
  Mori-S\'{a}nchez, and Yang}}]{Cohen12}
\bibinfo{author}{\bibfnamefont{A.~J.} \bibnamefont{Cohen}},
  \bibinfo{author}{\bibfnamefont{P.}~\bibnamefont{Mori-S\'{a}nchez}},
  \bibnamefont{and} \bibinfo{author}{\bibfnamefont{W.}~\bibnamefont{Yang}},
  \bibinfo{journal}{Chem. Rev.} \textbf{\bibinfo{volume}{112}},
  \bibinfo{pages}{289} (\bibinfo{year}{2012}).

\bibitem[{\citenamefont{K\"{u}mmel and Perdew}(2003)}]{KumPer03a}
\bibinfo{author}{\bibfnamefont{S.}~\bibnamefont{K\"{u}mmel}} \bibnamefont{and}
  \bibinfo{author}{\bibfnamefont{J.}~\bibnamefont{Perdew}},
  \bibinfo{journal}{Phys. Rev. Lett.} \textbf{\bibinfo{volume}{90}},
  \bibinfo{pages}{043004} (\bibinfo{year}{2003}).

\bibitem[{\citenamefont{K\"ummel and Perdew}(2003)}]{KumPer03b}
\bibinfo{author}{\bibfnamefont{S.}~\bibnamefont{K\"ummel}} \bibnamefont{and}
  \bibinfo{author}{\bibfnamefont{J.~P.} \bibnamefont{Perdew}},
  \bibinfo{journal}{Phys. Rev. B} \textbf{\bibinfo{volume}{68}},
  \bibinfo{pages}{035103} (\bibinfo{year}{2003}).

\bibitem[{\citenamefont{K\"ummel and Kronik}(2008)}]{KueKronik08}
\bibinfo{author}{\bibfnamefont{S.}~\bibnamefont{K\"ummel}} \bibnamefont{and}
  \bibinfo{author}{\bibfnamefont{L.}~\bibnamefont{Kronik}},
  \bibinfo{journal}{Rev. Mod. Phys.} \textbf{\bibinfo{volume}{80}},
  \bibinfo{pages}{3} (\bibinfo{year}{2008}).

\end{thebibliography}

\end{document}


\title{Supplemental Material to: \emph{"Elimination of the asymptotic fractional dissociation problem in Kohn-Sham density functional theory using the ensemble-generalization approach"}}

\author{Eli Kraisler}
\affiliation{Department of Materials and Interfaces, Weizmann Institute of Science, Rehovoth 76100,Israel}

\author{Leeor Kronik}
\affiliation{Department of Materials and Interfaces, Weizmann Institute of Science, Rehovoth 76100,Israel}

\date{\today}

\maketitle

\textbf{Contents}

\begin{enumerate}[I.]
  \item Energy curves $E(N)$ for the atoms H, Li, C, and F.
  \item Energy curves $E(q)$ for infinitely separated molecules.
  \item Approaches to negatively charged ions.
\end{enumerate}

\section{Energy curves $\maybebm{E(N)}$ for the atoms H, L$\textrm{i}$, C, and F}

\begin{figure}[H]
  \centering
  \includegraphics[trim= 0mm -5mm 0mm 0mm,scale=0.6]{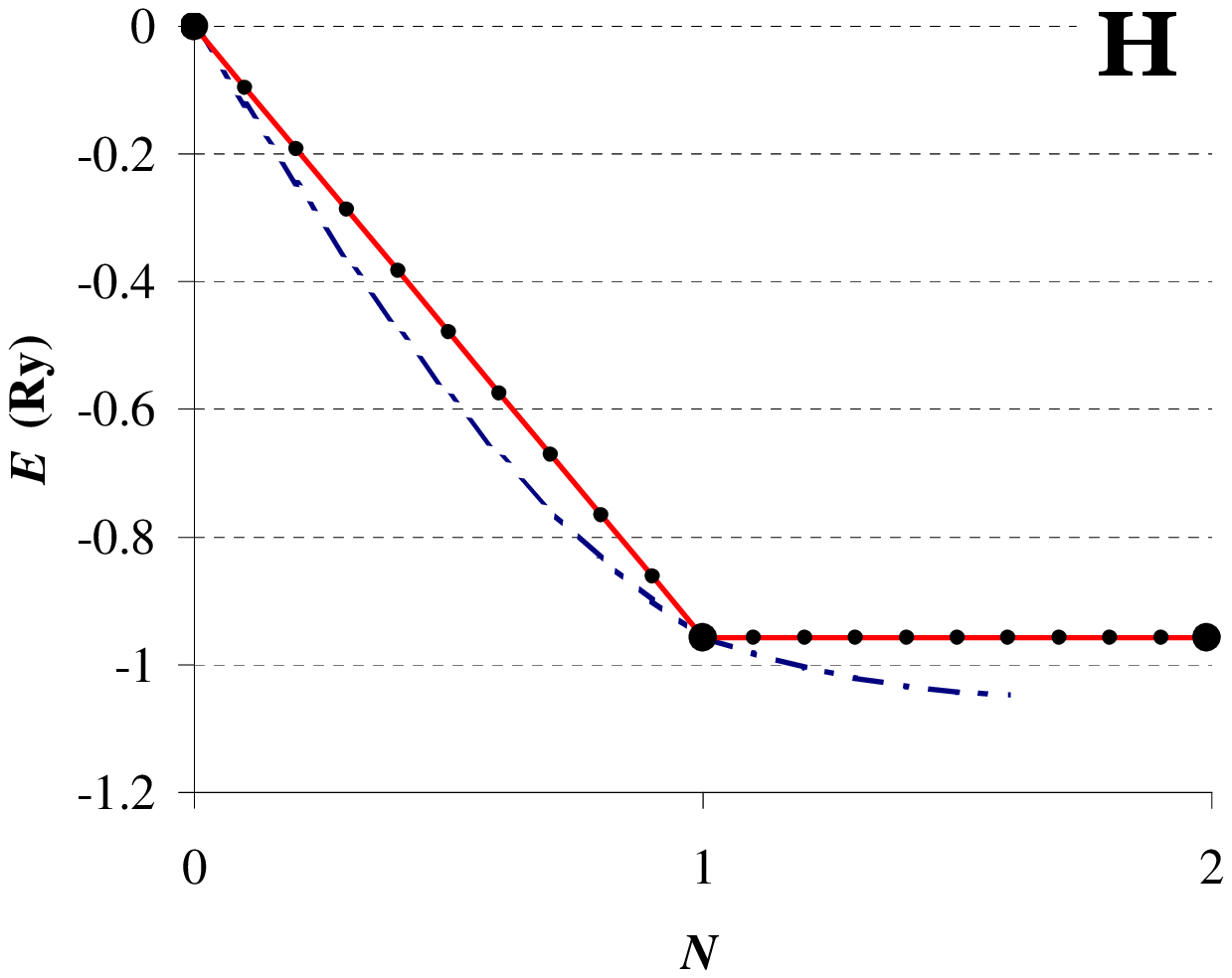}
  \includegraphics[trim=-5mm -5mm 0mm 0mm,scale=0.6]{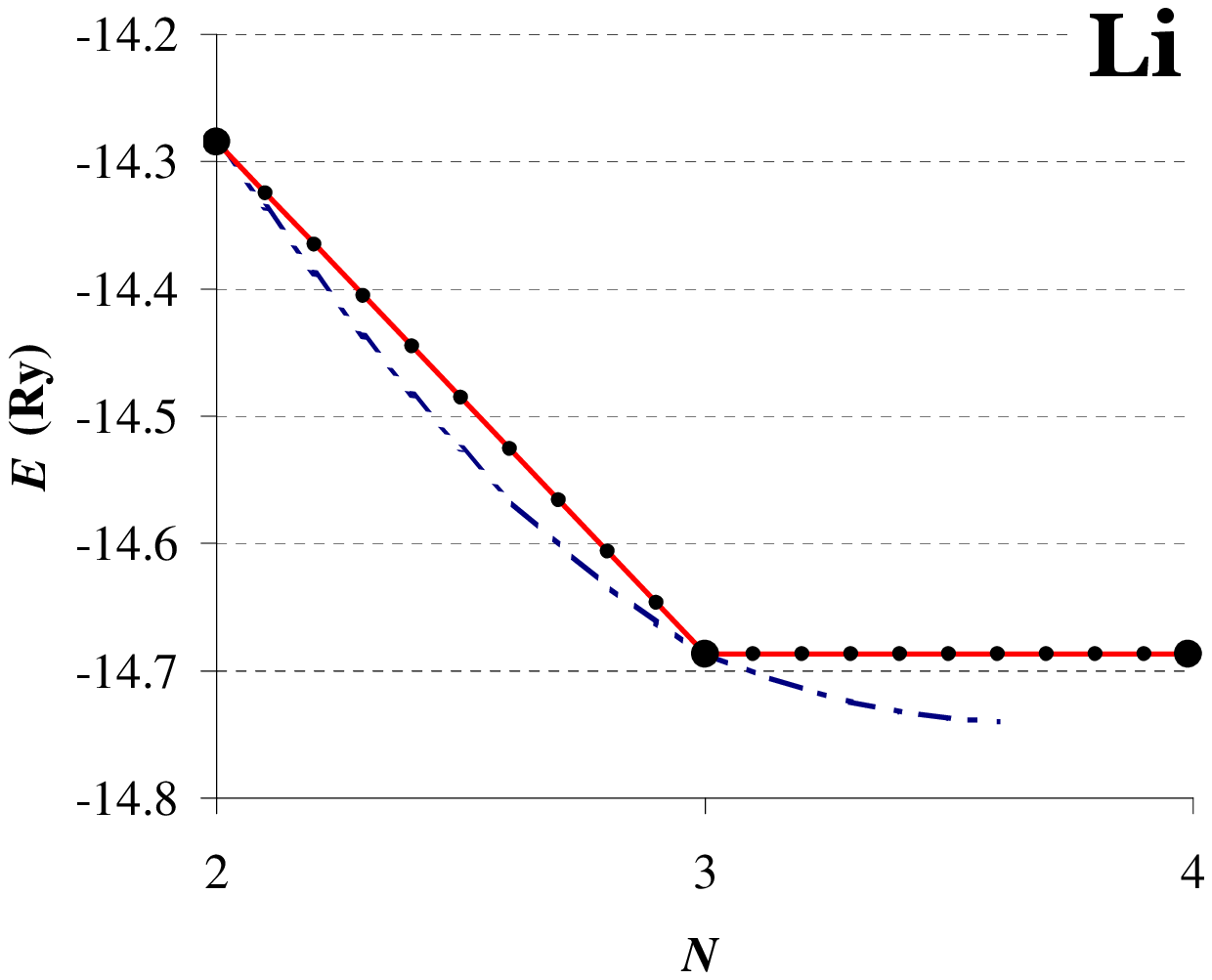}
  \includegraphics[trim= 0mm  0mm 0mm 0mm,scale=0.6]{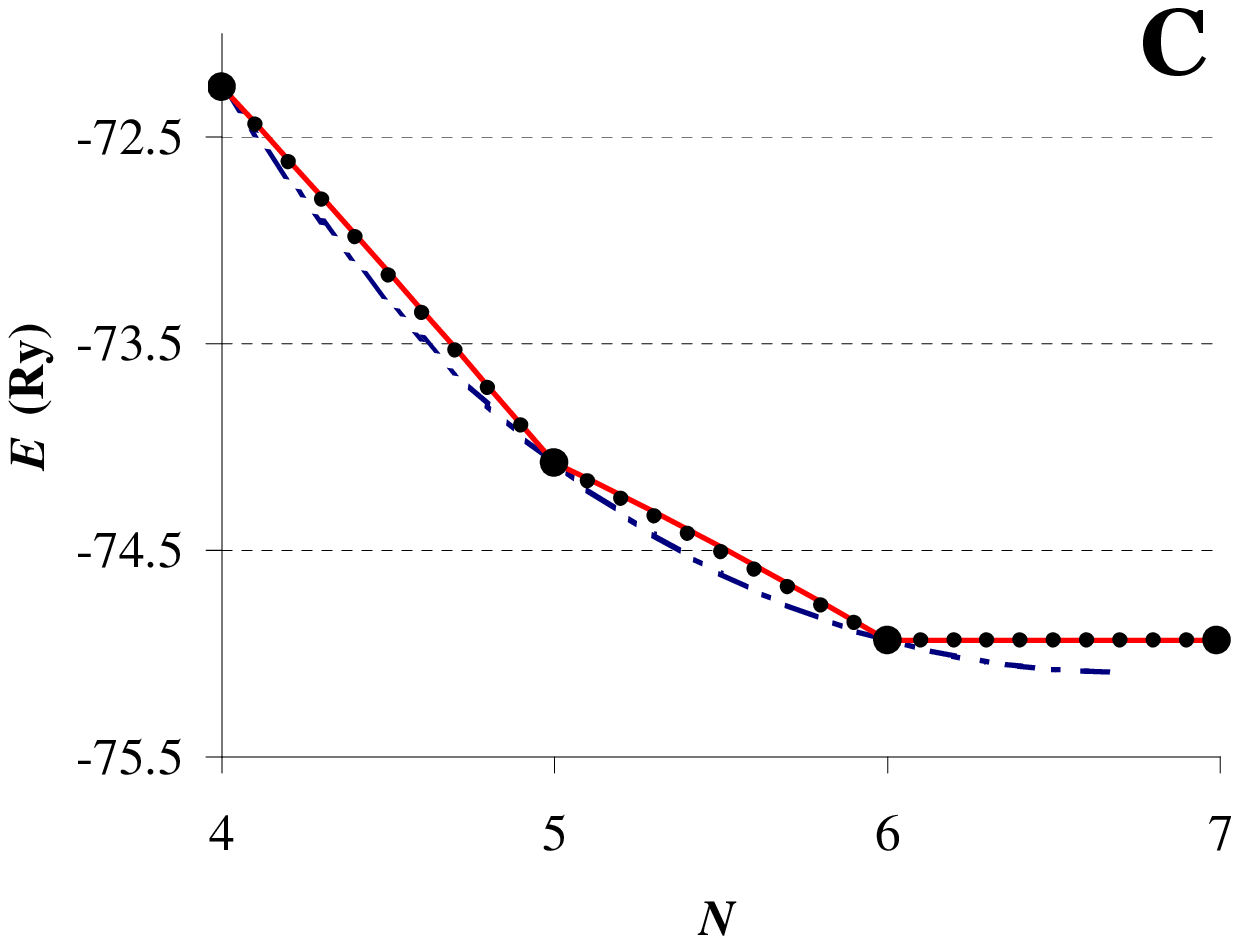}
  \includegraphics[trim=-5mm  0mm 0mm 0mm,scale=0.6]{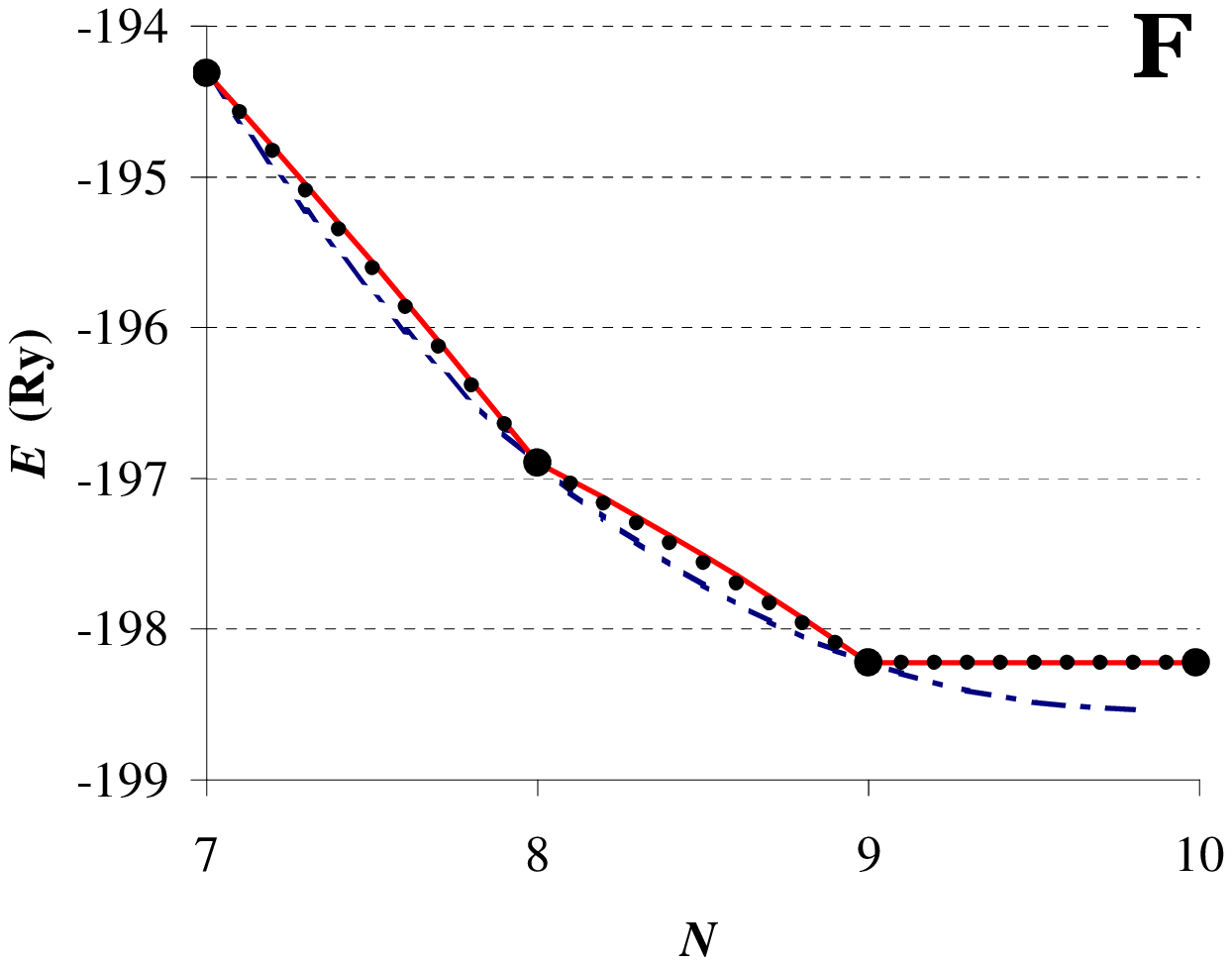}
  \caption{Energy as a function of the number of electrons, for the atoms H, Li, C, and F, computed with LSDA (dash-dotted line) and eLSDA (solid line), compared to the expected piecewise-linear (dotted line) behavior (the latter is obtained by linear interpolation of the eLSDA energies at integer electron values).}
  \label{fig.E_N} 
\end{figure}

Figure~\ref{fig.E_N} presents the dependence of the energy, $E(N)$, on the number of electrons, $N$, obtained for various atoms, using the LSDA and eLSDA. The results are compared to the piecewise-linear behavior obtained by linear interpolation between eLSDA values at integer $N$. Data from these curves were used to obtain the energy curves of infinitely separated molecules, as shown in Sec.~\ref{sec.E.mol} below. 

From Fig.~\ref{fig.E_N} one can clearly see that the eLSDA is much closer to piecewise-linearity in all cases. As mentioned in the main text, LSDA results are presented only if the obtained system is bound, which is not always the case for (fractionally) negatively charged ions. 
For eLSDA, where negative ions were never obtained as bound, their energy was taken to be equal to that of a neutral system (plus a fraction of an electron at infinity, whose contribution is zero)~\cite{Tozer07,Teale08}. An alternative approach to negative ion calculations is discussed in Sec.~\ref{sec.neg.ions} below. Note that in the context of the fractional dissociation problem both approaches lead to the same conclusions.

\section{Energy curves $\maybebm{E(q)}$ for infinitely separated molecules} \label{sec.E.mol}

\begin{figure}[H]
  \centering
  \includegraphics[trim= 0mm -5mm 0mm 0mm,scale=0.6]{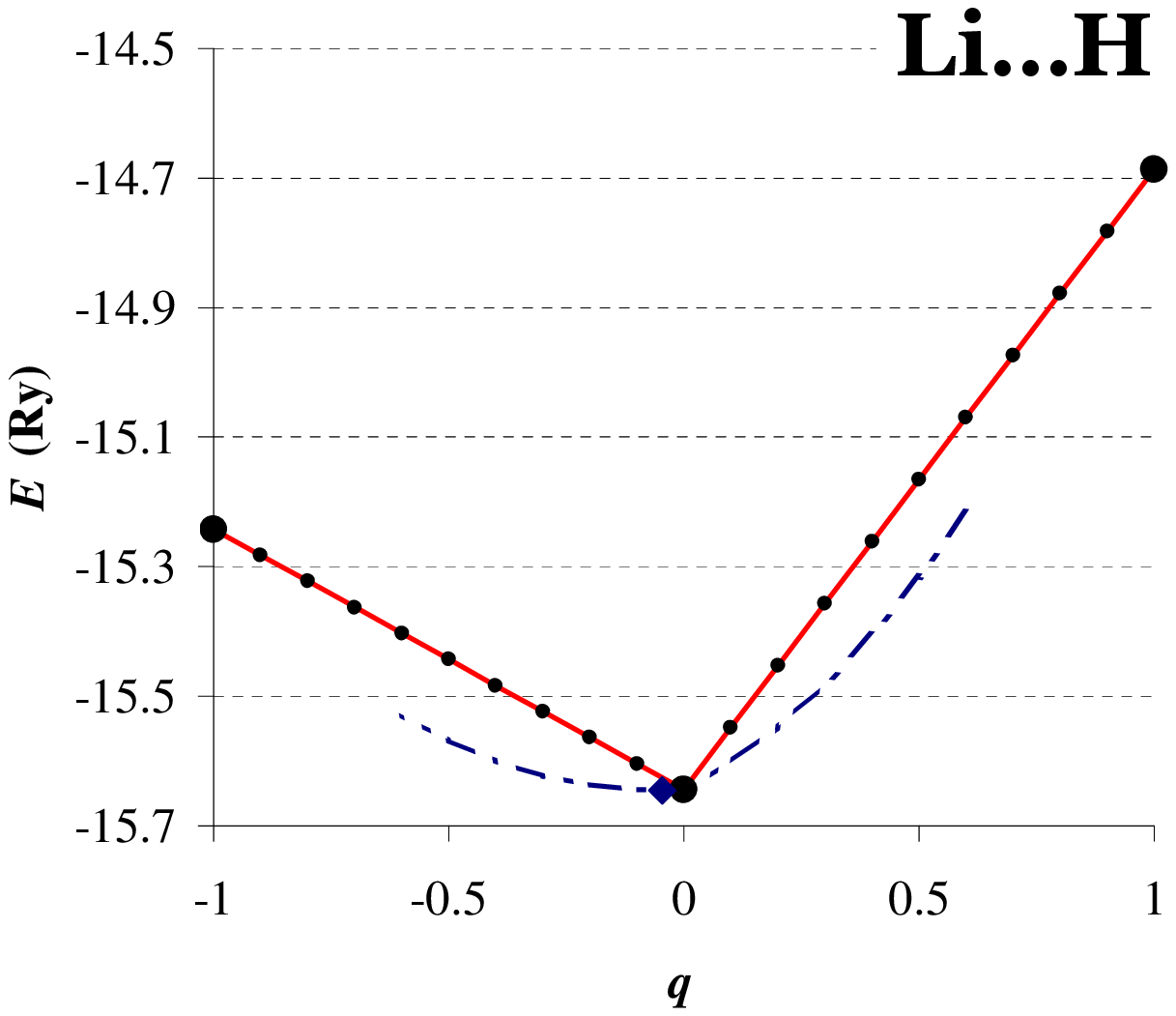}
  \includegraphics[trim=-5mm -5mm 0mm 0mm,scale=0.6]{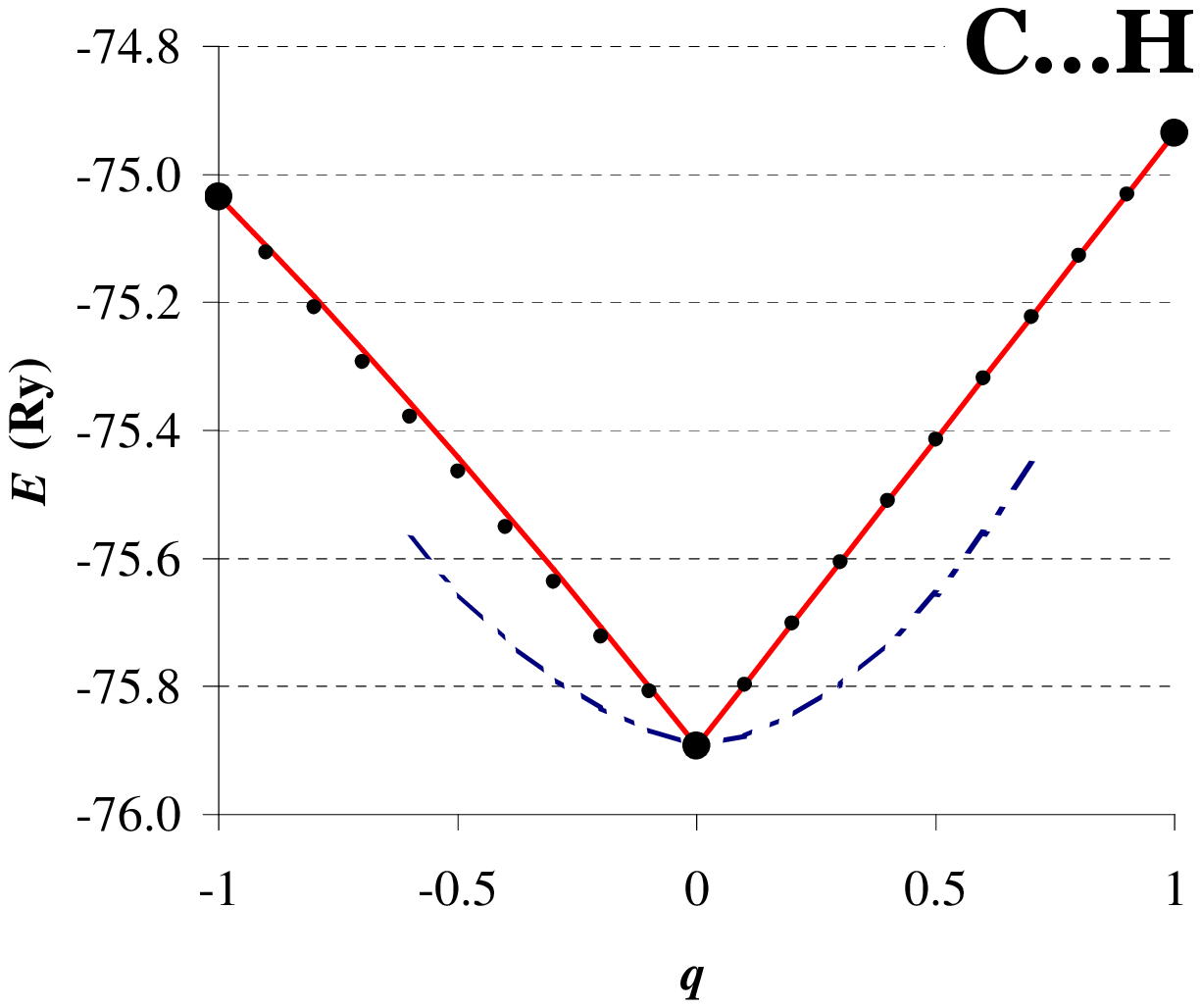}
  \includegraphics[trim= 0mm  0mm 0mm 0mm,scale=0.6]{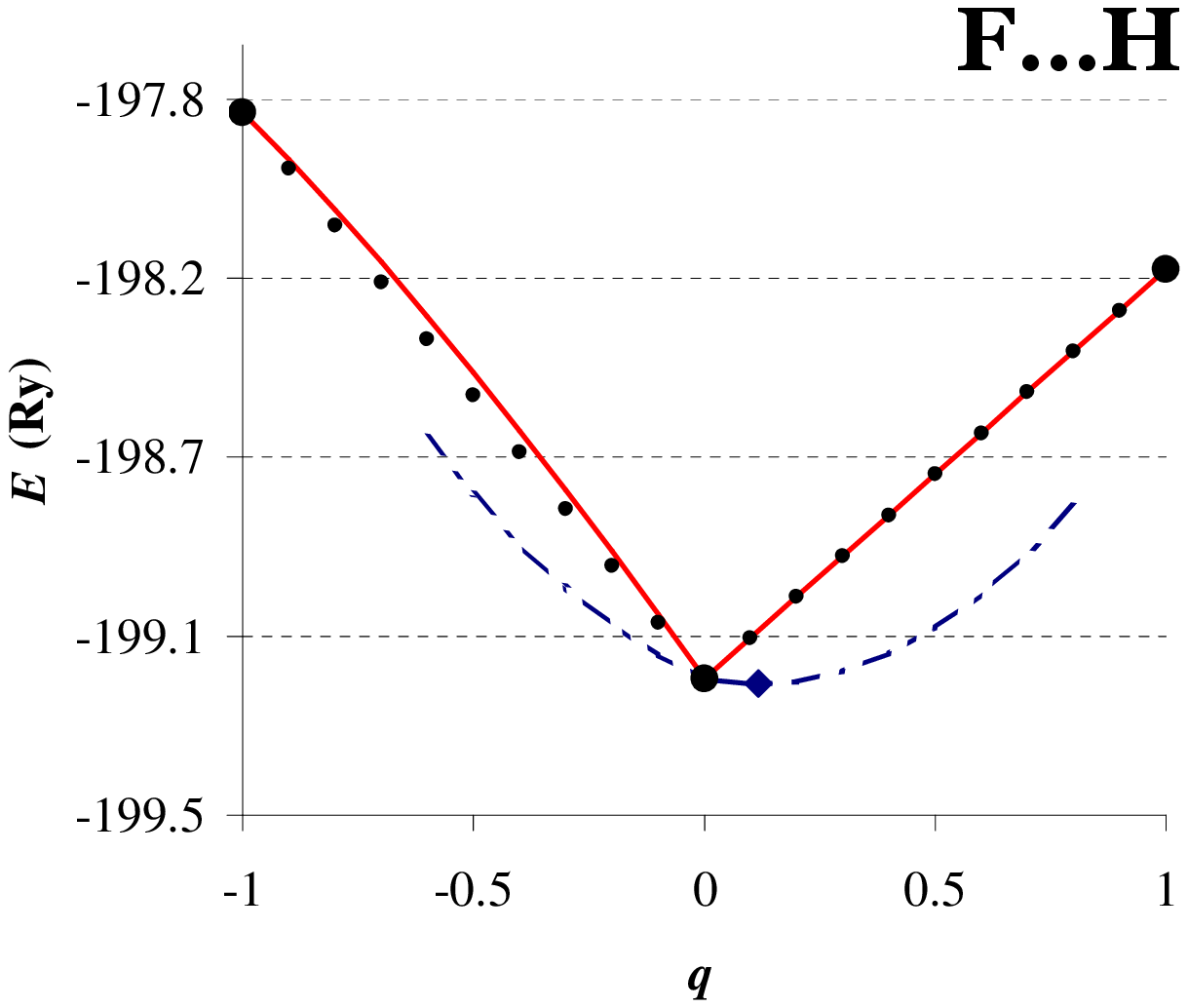}
  \includegraphics[trim=-5mm  0mm 0mm 0mm,scale=0.6]{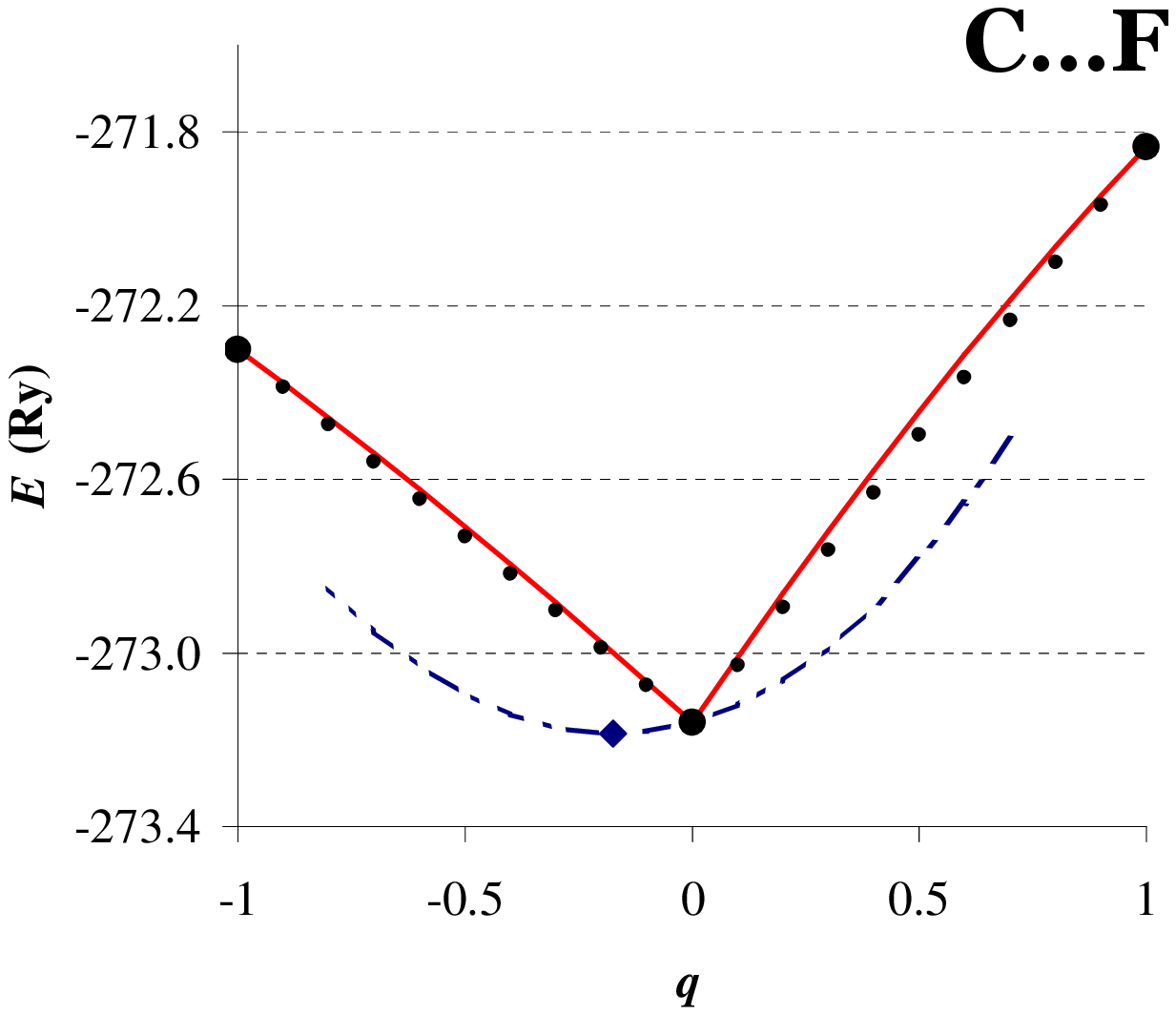}
  
  \caption{Energy as a function of $q$ for the neutral, heteroatomic molecules Li...H, C...H, F...H, and C...F, computed with LSDA (dash-dotted line) and eLSDA (solid line), compared to the expected piecewise-linear (dotted line) behavior (the latter is obtained by linear interpolation of the eLSDA energies at integer electron values).}
  \label{fig.E_q_AB} 
\end{figure}

\begin{figure}[H]
  \centering
  \includegraphics[trim= 0mm -5mm 0mm 0mm,scale=0.6]{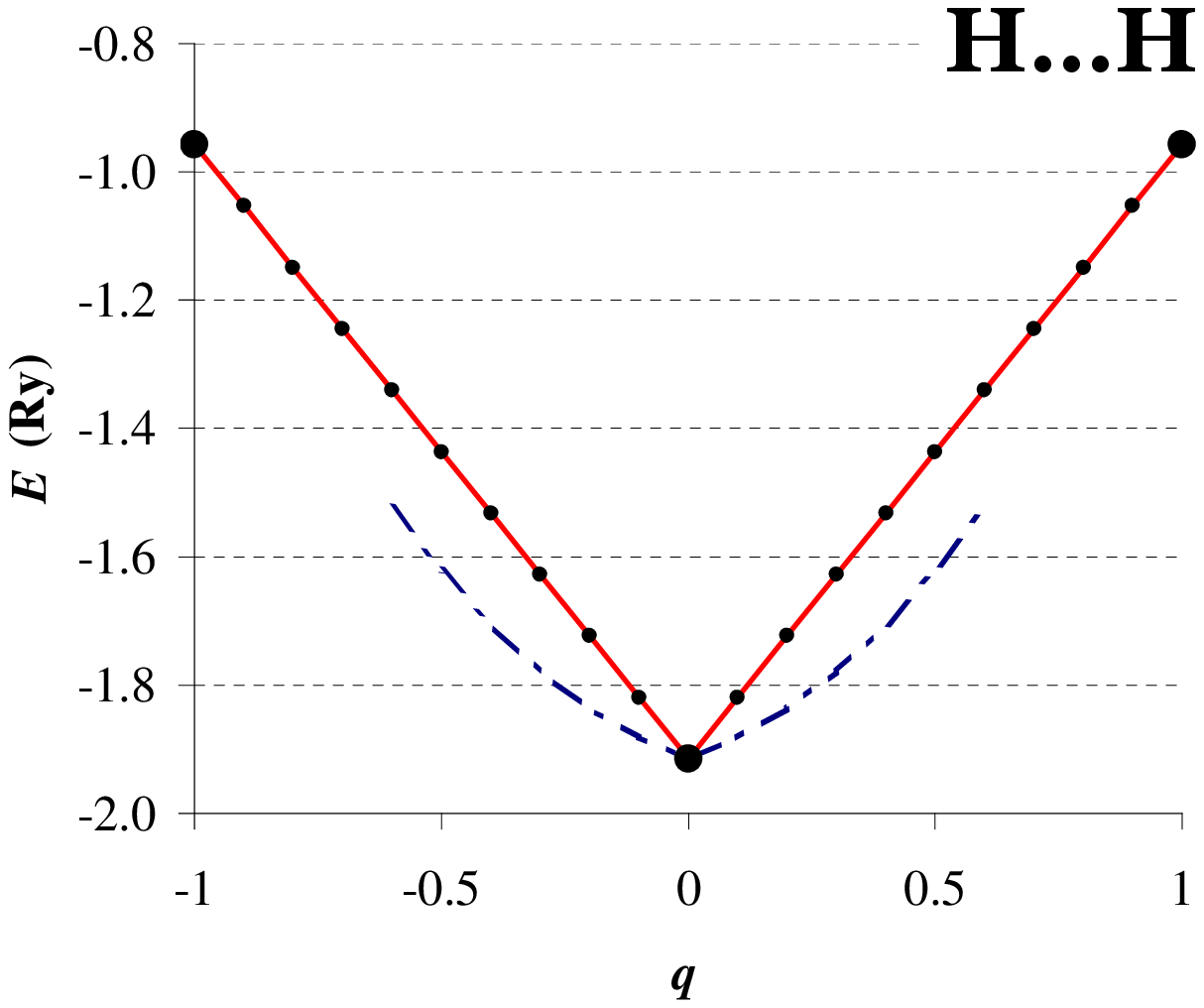}
  \includegraphics[trim=-5mm -5mm 0mm 0mm,scale=0.6]{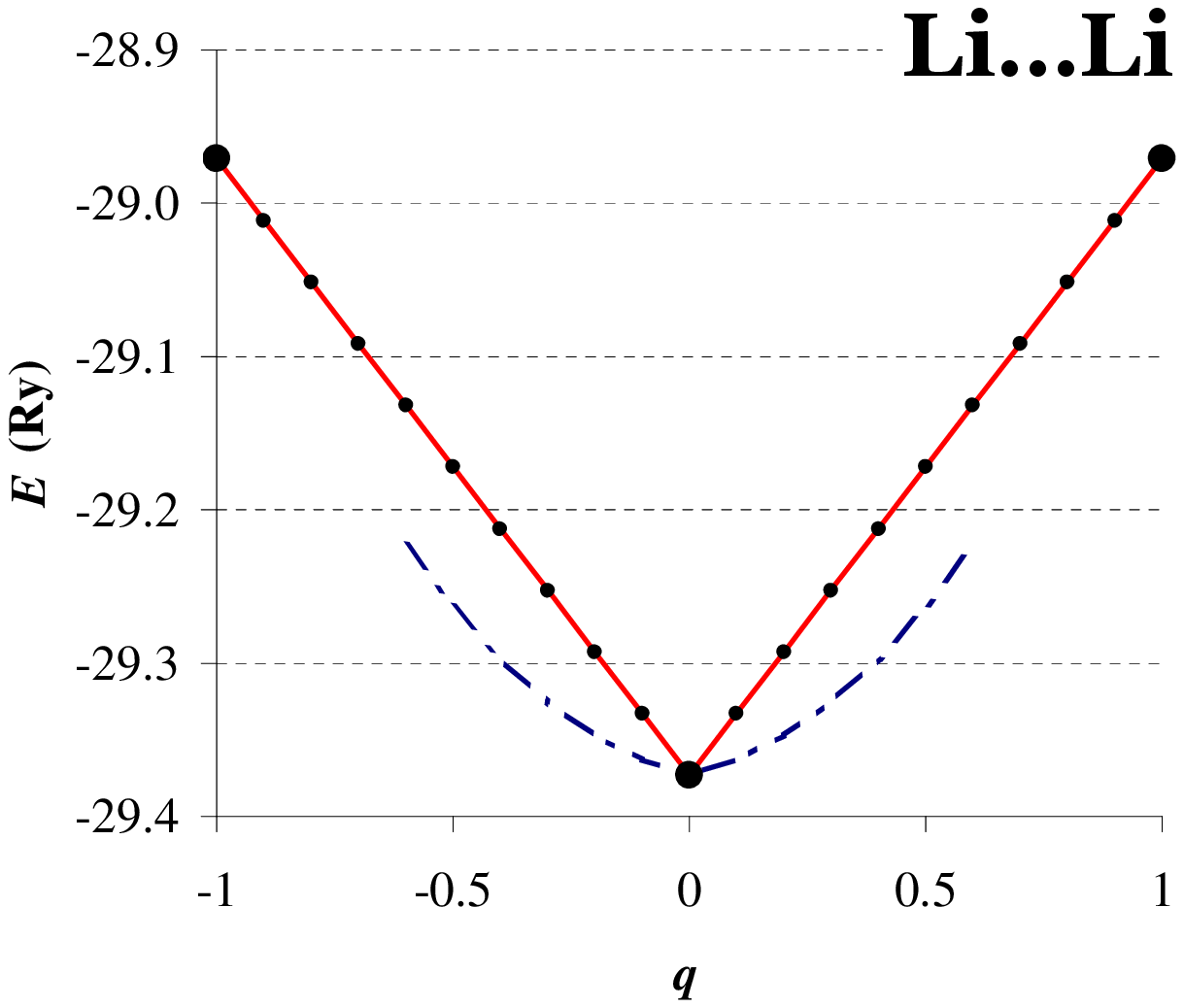}
  \includegraphics[trim= 0mm  0mm 0mm 0mm,scale=0.6]{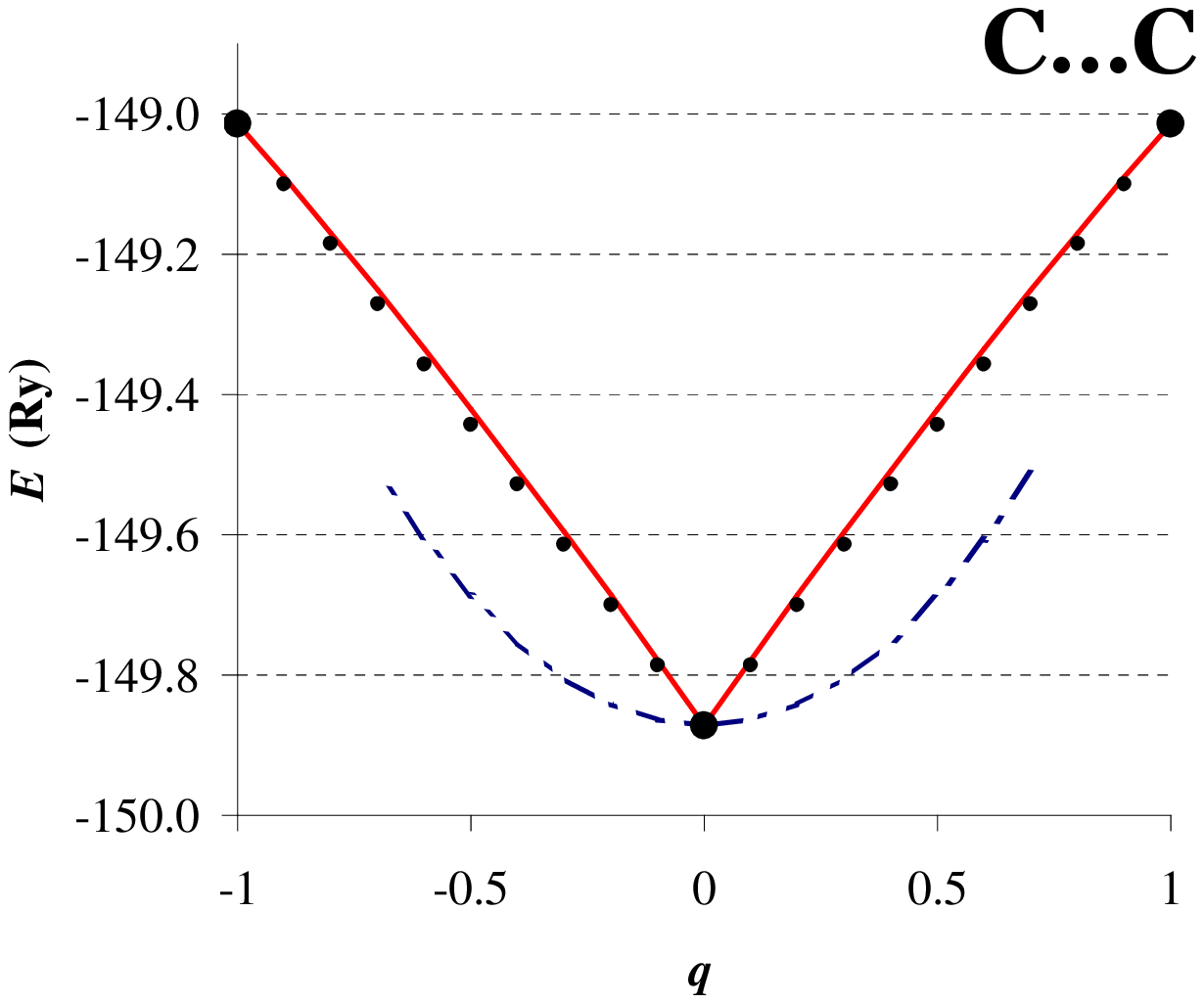}
  \includegraphics[trim=-5mm  0mm 0mm 0mm,scale=0.6]{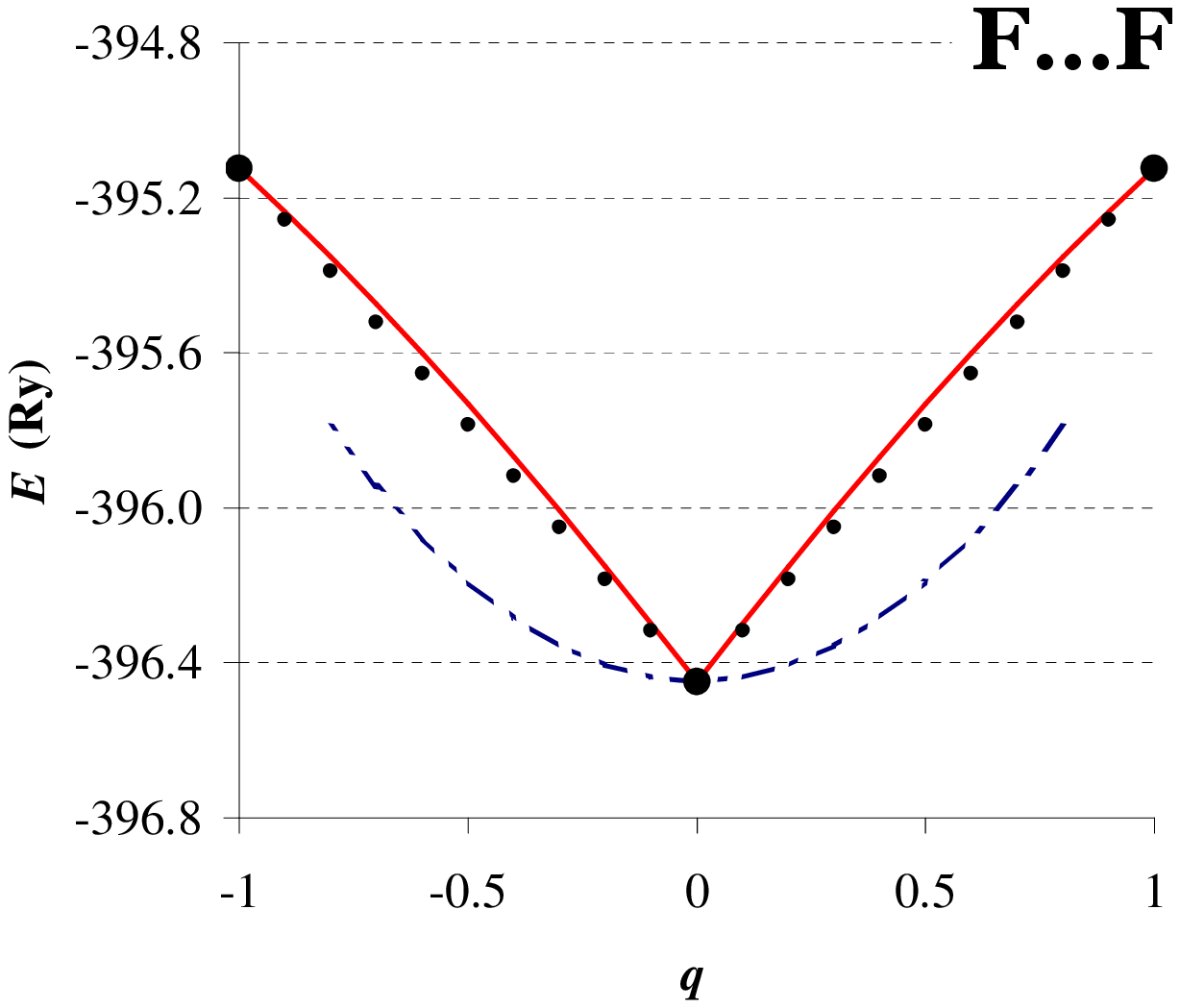}
  
  \caption{Energy as a function of $q$ for the neutral, homoatomic molecules H...H, Li...Li, C...C, and F...F, computed with LSDA (dash-dotted line) and eLSDA (solid line), compared to the expected piecewise-linear (dotted line) behavior (the latter is obtained by linear interpolation of the eLSDA energies at integer electron values).}
  \label{fig.E_q_AA} 
\end{figure}

\begin{figure}[H]
  \centering
  \includegraphics[trim= 0mm -5mm 0mm 0mm,scale=0.6]{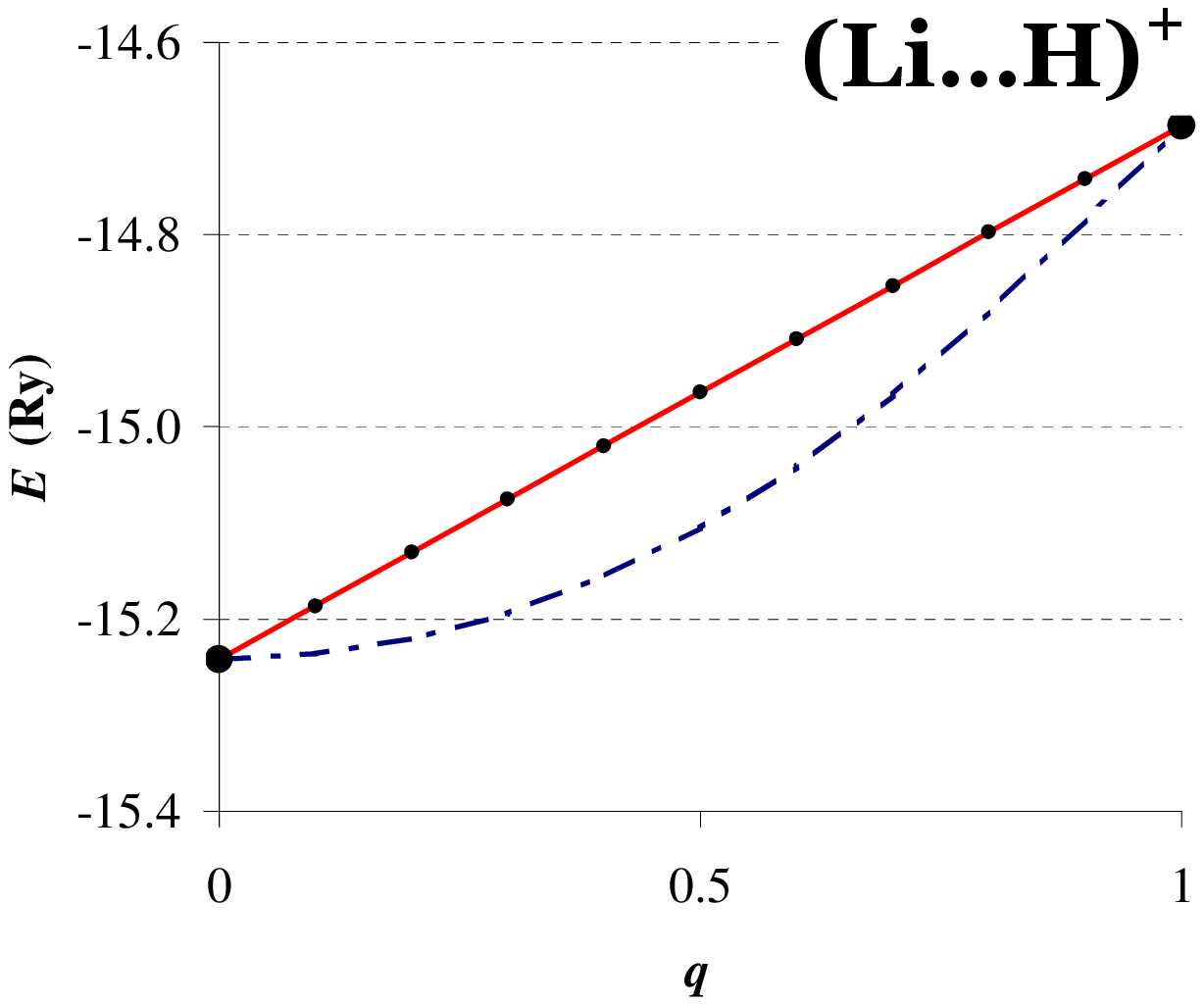}
  \includegraphics[trim=-5mm -5mm 0mm 0mm,scale=0.6]{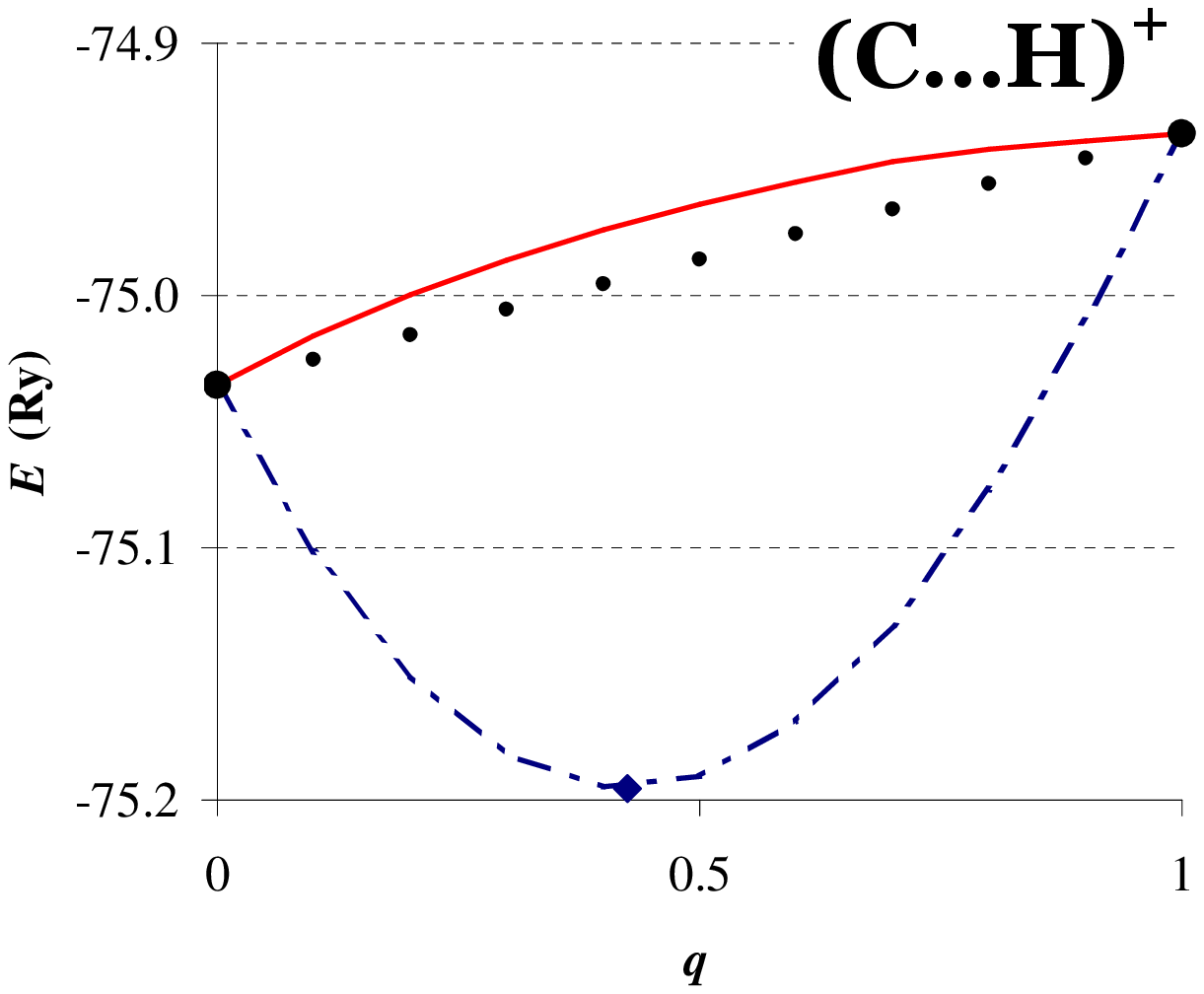}
  \includegraphics[trim= 0mm -5mm 0mm 0mm,scale=0.6]{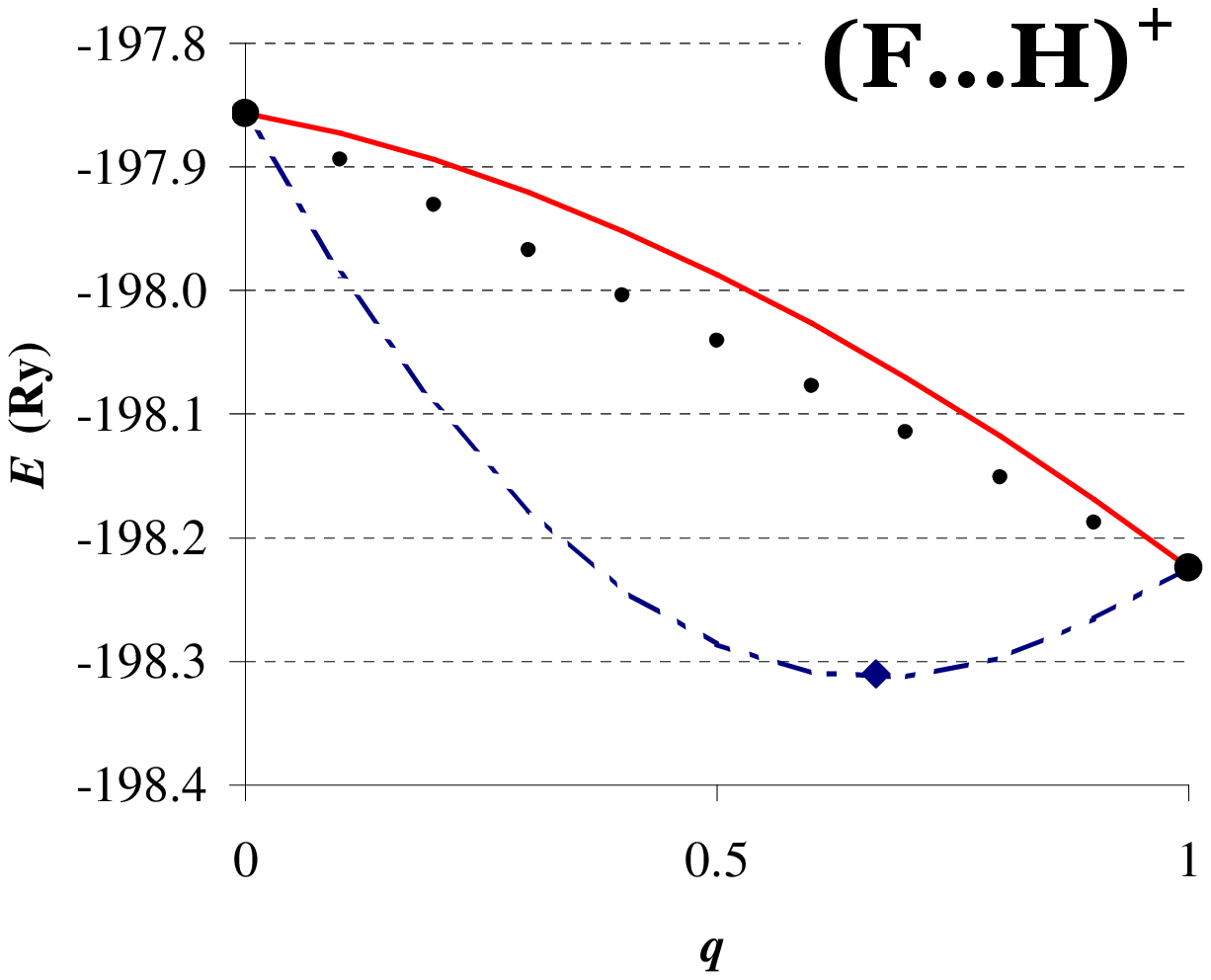}
  \includegraphics[trim=-5mm -5mm 0mm 0mm,scale=0.6]{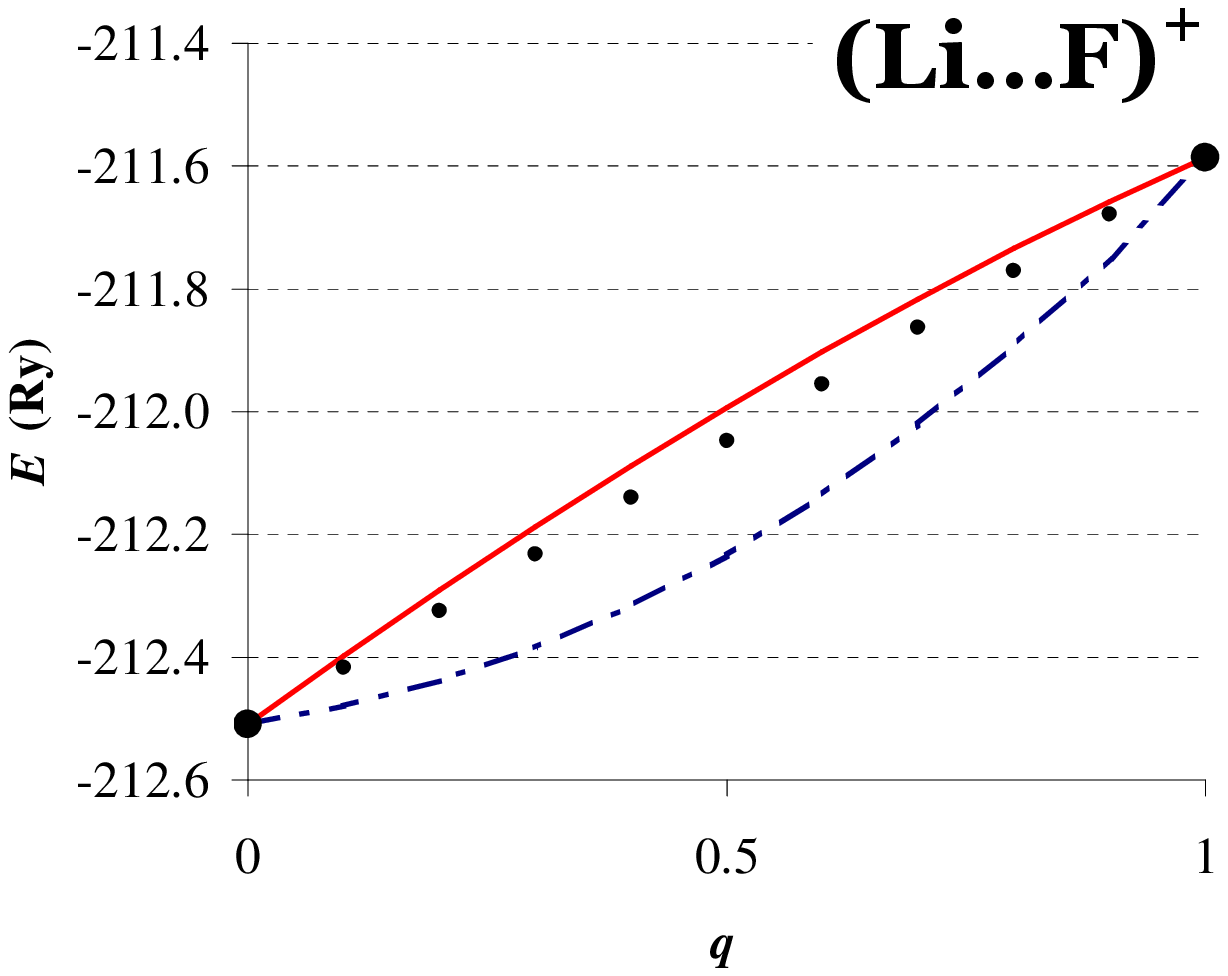}
  \includegraphics[trim= 0mm  0mm 0mm 0mm,scale=0.6]{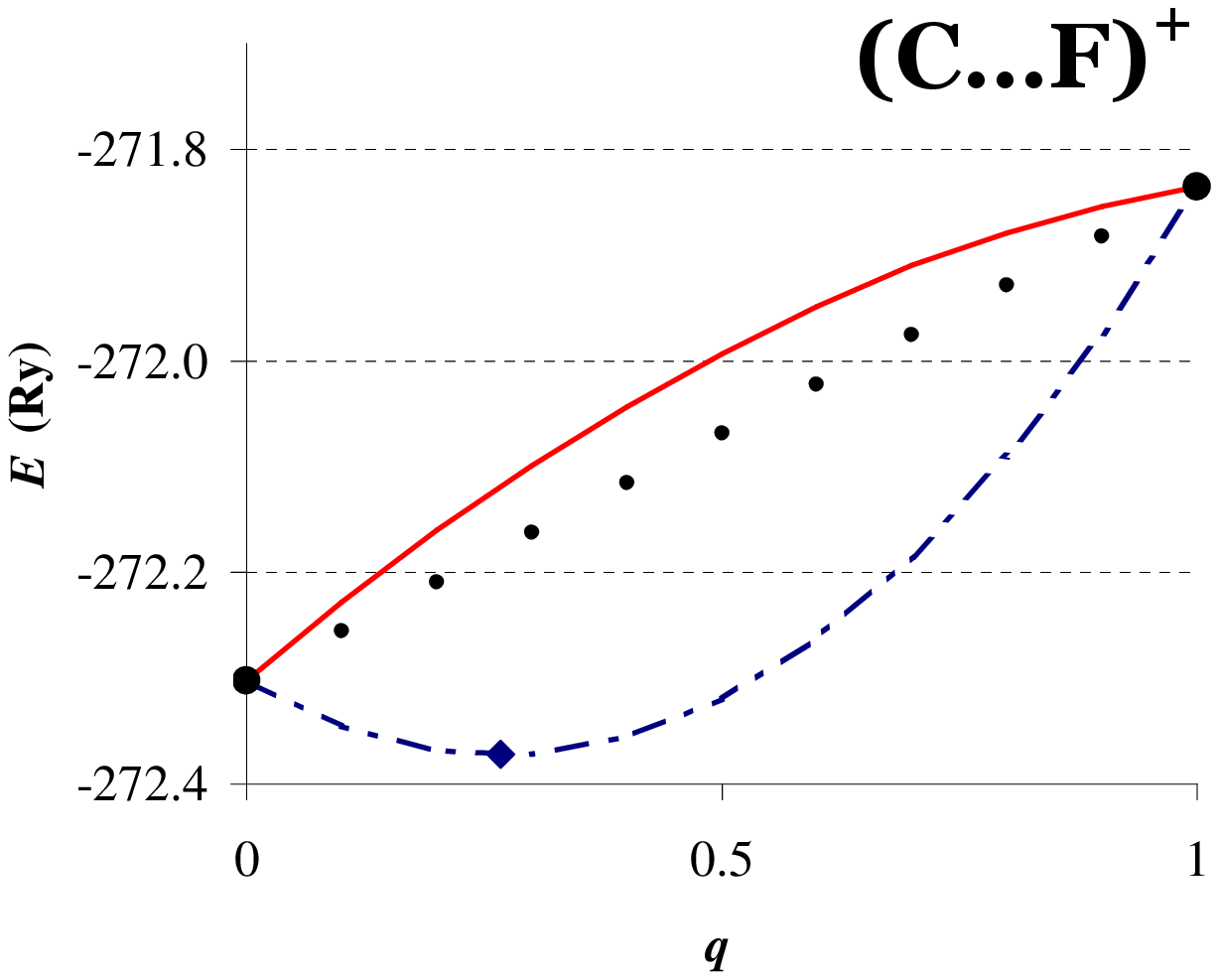}
  
  \caption{Energy as a function of $q$ for the singly ionized, heteroatomic molecules (Li...H)$^+$, (C...H)$^+$, (F...H)$^+$, (Li...F)$^+$, and (C...F)$^+$, computed with LSDA (dash-dotted line) and eLSDA (solid line), compared to the expected piecewise-linear (dotted line) behavior (the latter is obtained by linear interpolation of the eLSDA energies at integer electron values).}
  \label{fig.E_q_AB_ions} 
\end{figure}

\begin{figure}[H]
  \centering
  \includegraphics[trim= 0mm -5mm 0mm 0mm,scale=0.6]{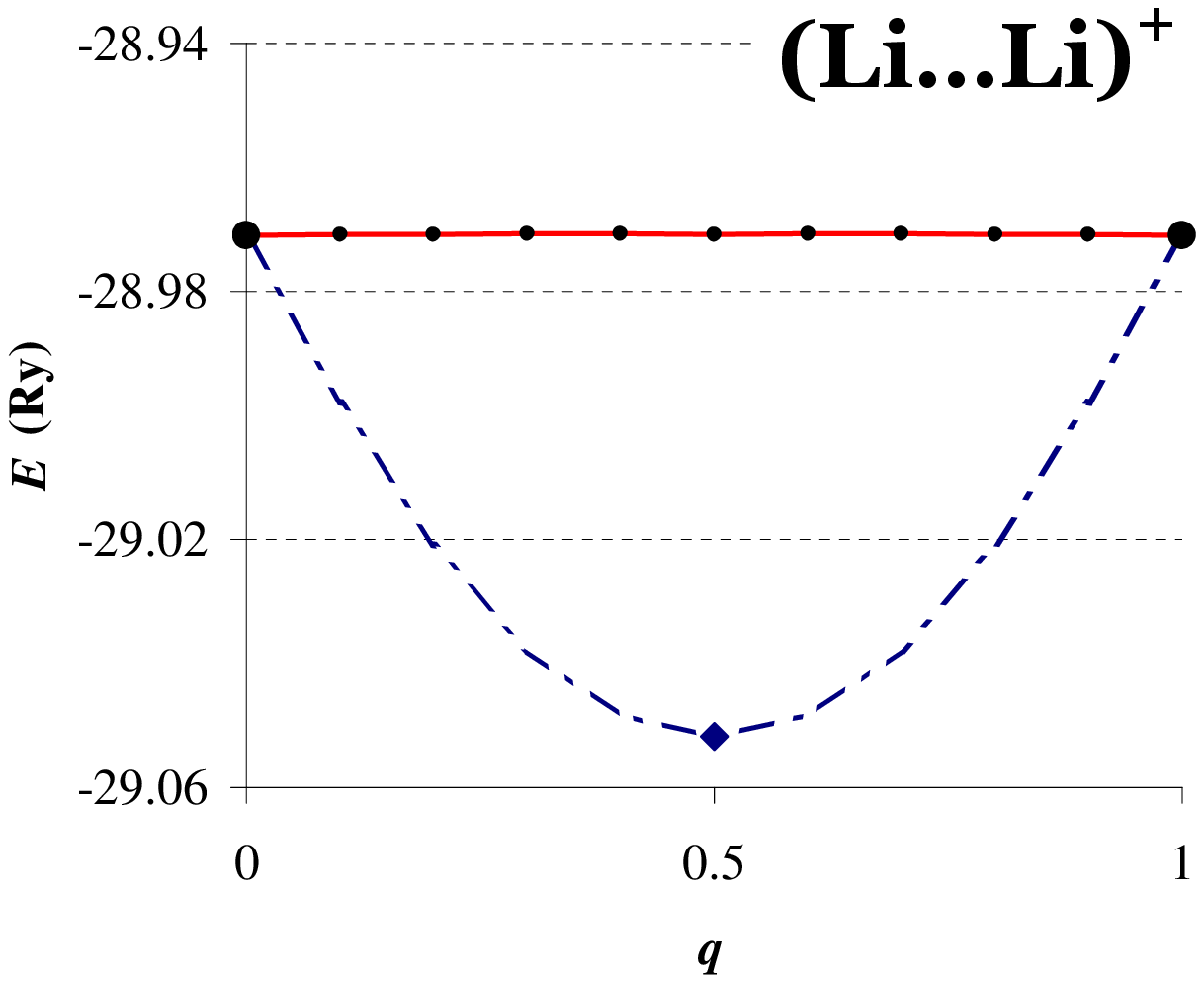}
  \includegraphics[trim=-5mm -5mm 0mm 0mm,scale=0.6]{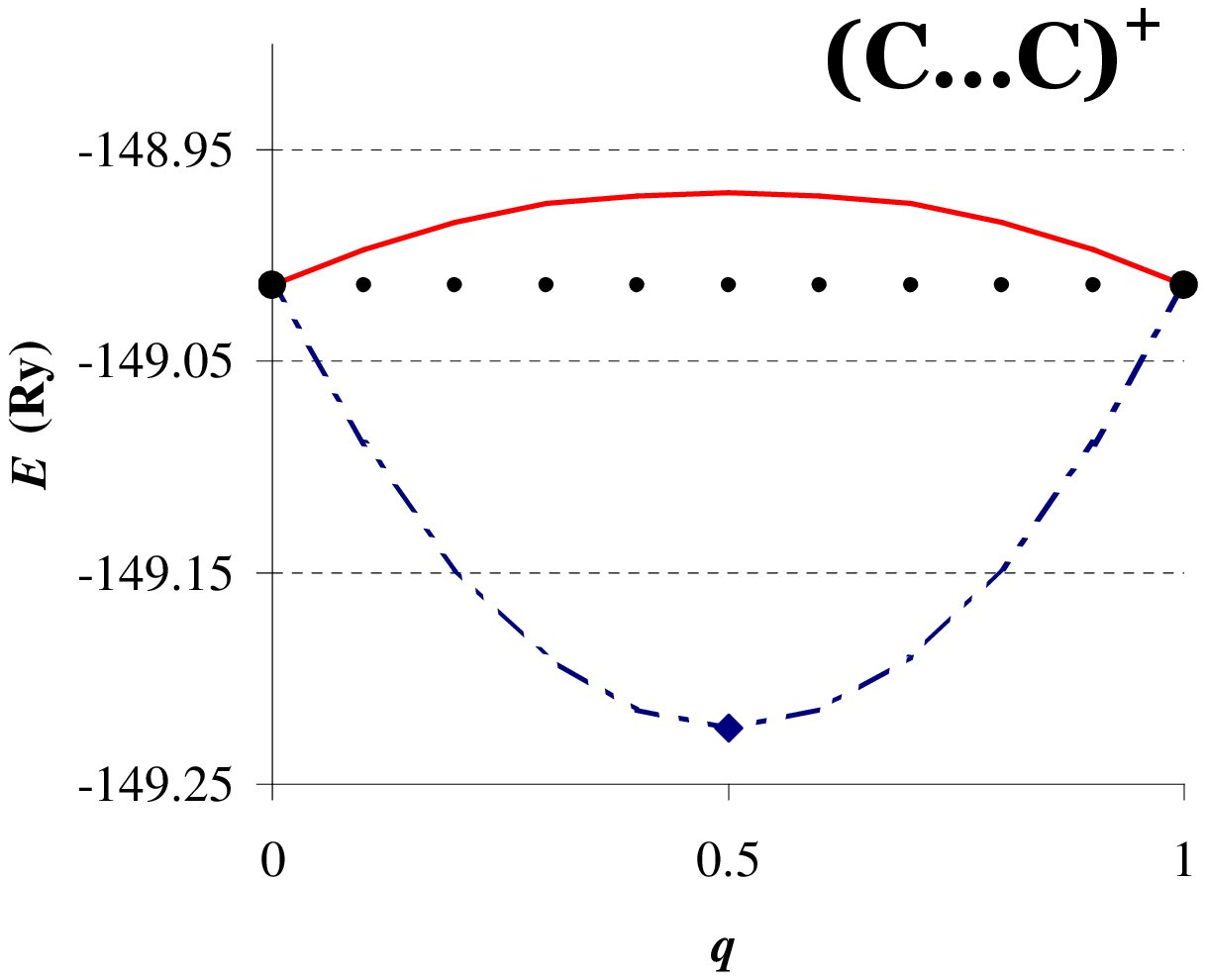}
  
  \caption{Energy as a function of $q$ for the singly ionized, homoatomic molecules (Li...Li)$^+$ and (C...C)$^+$, computed with LSDA (dash-dotted line) and eLSDA (solid line), compared to the expected piecewise-linear (dotted line) behavior (the latter is obtained by linear interpolation of the eLSDA energies at integer electron values).}
  \label{fig.E_q_AA_ions} 
\end{figure}

\begin{figure}[H]
  \centering
\includegraphics[trim=0mm 0mm 0mm 0mm,scale=0.6]{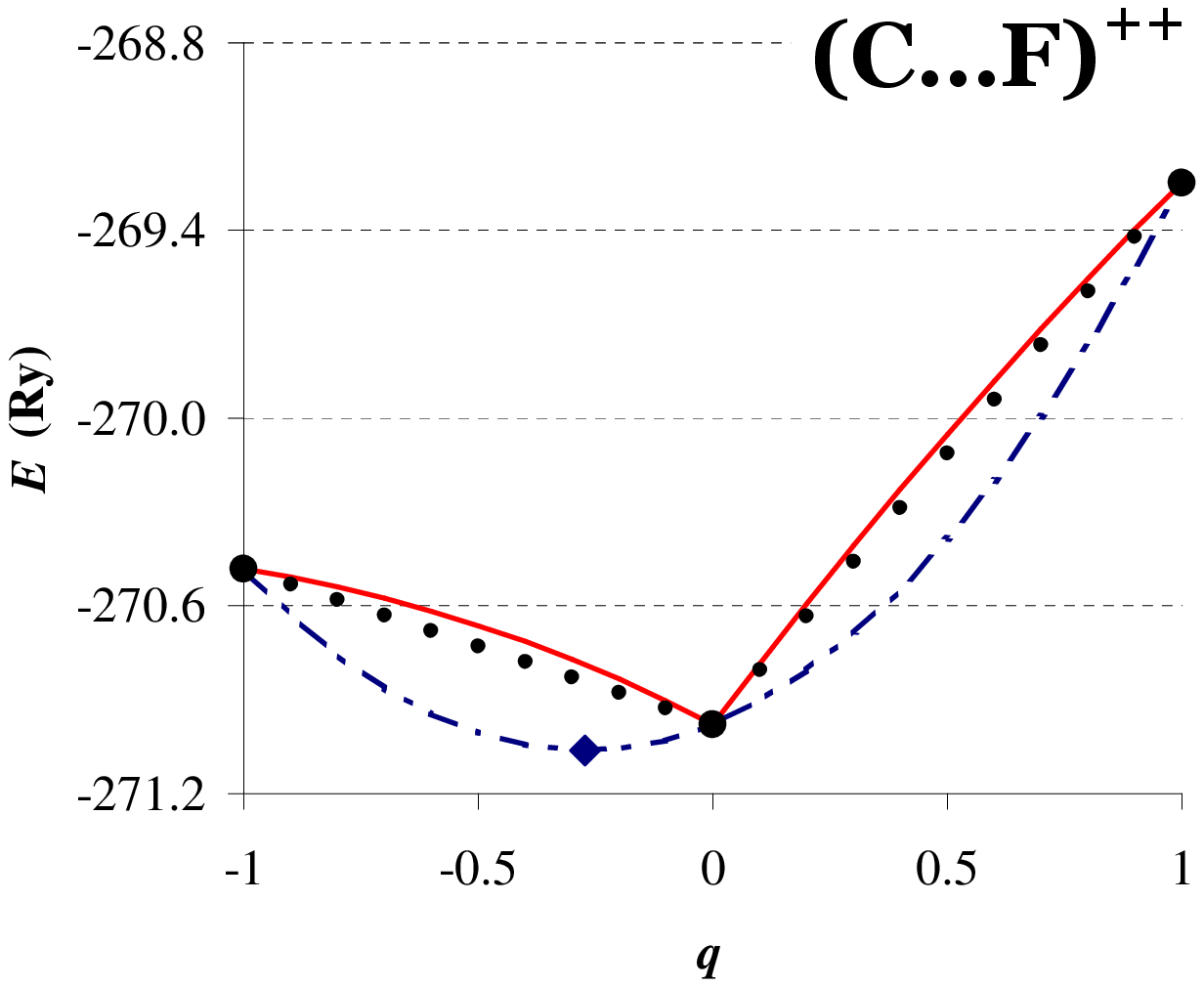}
  
  \caption{Energy as a function of $q$ for the doubly ionized molecule (C...F)$^{++}$, computed with LSDA (dash-dotted line) and eLSDA (solid line), compared to the expected piecewise-linear (dotted line) behavior (the latter is obtained by linear interpolation of the eLSDA energies at integer electron values).}
  \label{fig.E_q_CF++} 
\end{figure}

Using the data from atomic calculations, presented in Fig.~\ref{fig.E_N}, we can obtain energy curves $E_{A...B}(q)$ for infinitely separated molecules, using Eq.~(\EqEq) from the main text. $q$ is the amount of negative charge transferred from atom $B$ to atom $A$.
Figure~\ref{fig.E_q_AB} presents the energy curves for neutral, heteroatomic molecules. Spurious mimina obtained with the LSDA are marked by rhombi. We readily see that these minima are absent from eLSDA calculations in all cases examined. Figure~\ref{fig.E_q_AA} presents the energy curves for neutral, homoatomic molecules. For these systems, no spurious minima are expected, due to symmetry considerations. However, we see that eLSDA is much closer to piecewise-linearity. Figure~\ref{fig.E_q_AB_ions} presents the energy curves for singly ionized, heteroatomic molecules and Fig.~\ref{fig.E_q_AA_ions} for singly ionized, homoatomic molecules. Spurious mimina obtained with the LSDA are marked by rhombi. We again readily see that these minima are absent from eLSDA calculations in all cases examined. Finally, Fig.~\ref{fig.E_q_CF++} shown the energy curve for the (C...F)$^{++}$ molecule, which has been chosen in the main text to study the KS potential behavior.

\section{Approaches to negatively charged ions}\label{sec.neg.ions}

In order to describe an infinitely separated diatomic molecule, while transferring a fraction of an electron from one of its atom to the other one, we sometimes need to perform atomic calculations for ions with a negative net charge, i.e., when the (possibly fractional) number of electrons, $N$, is larger than the atomic number, $Z$. It is well-known that the LSDA and eLSDA may yield unbound solutions for such systems~\cite{Primer}. 

Principally, there are two approaches to address the aforementioned case of $N>Z$~\cite{Tozer07,Teale08}: (i) To state that since the negative ion is not bound, its energy, as well as the energy of an ion with any fractional $N$ larger than $Z$, equals the energy of the neutral atom plus the energy of (a fraction of) an electron at infinity. Since the contribution of the latter is zero, for $N>Z$ the energy $E(N)=E(Z)$, i.e., it is a horizontal line, independent of $N$; (ii) To obtain the energy of the negatively charged system from a DFT calculation (see, e.g.,~\cite{MoriS06,Vydrov07,Cohen12} and references therein). Here, however, we have to note that in cases where the system is not bound, calculations are normally not converged in terms of the basis set, or in our case of real-space calculations -- in terms of $\mu_{max}$ -- the maximal value of the prolate-spheroidal coordinate $\mu$, which plays here roughly the same role as the radial coordinate $r$ in spherical coordinates. Therefore, usually the grid of the neutral system is used for negative ions, as well, although it is formally not converged. 

In the current work, we chose approach (i). In this section we show that employment of approach (ii) to total-energy calculations of infinitely-separated molecules with LSDA and eLSDA, although yeilding results that are different quantitatively, leads to the same qualitative conclusions as approach (i): removal of the spurious energy minimum at fractional $q$ by eLSDA and strongly reduced deviation from piecewise linearity. For illustration we consider the F atom and the molecule Li...F.

\begin{figure}[H]
\centering
  \includegraphics[scale=0.6]{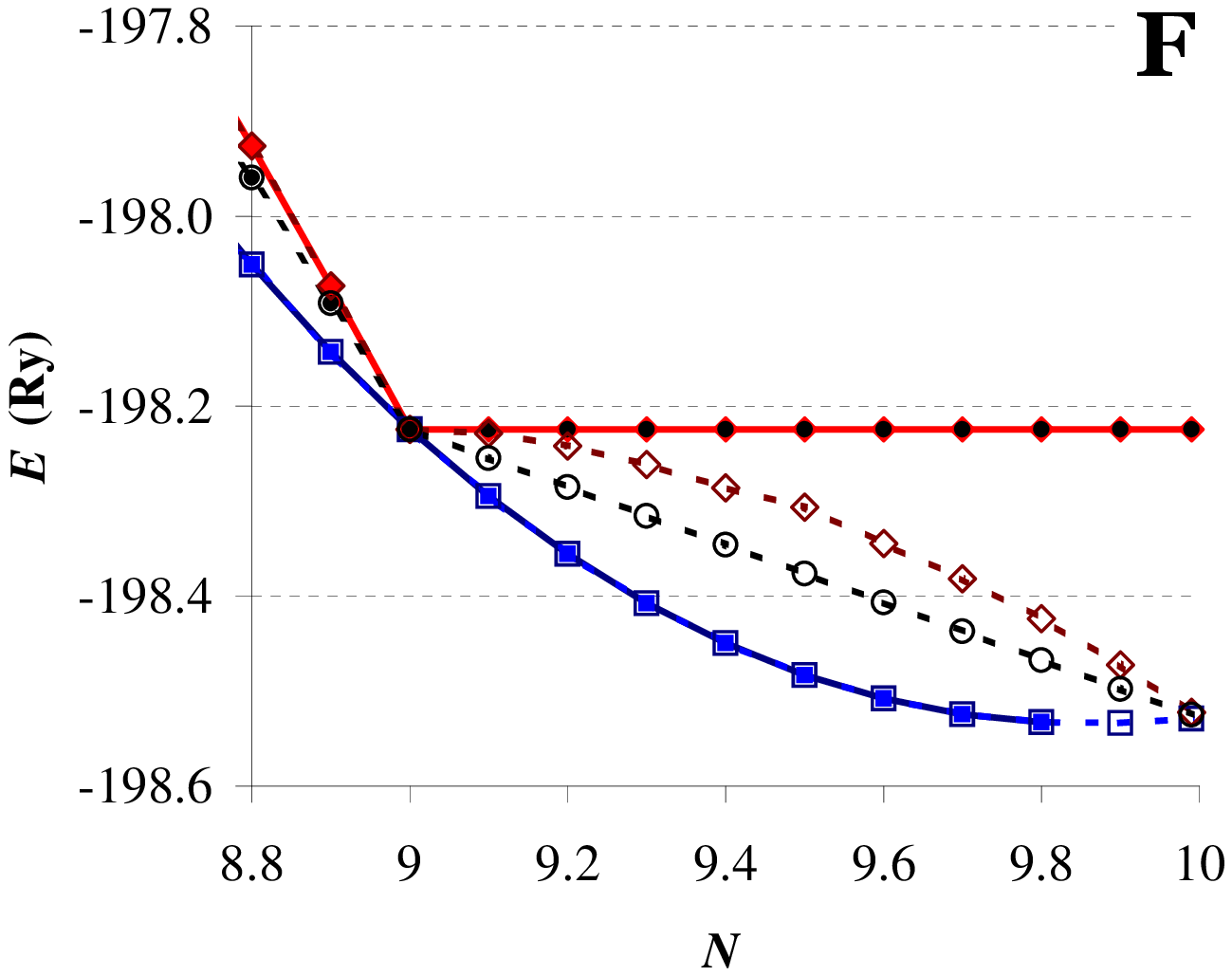}
  \caption{Energy as a function of $N$ for the F atom, obtained using: approach (i) -- with LSDA (blue, closed squares, solid line) and eLSDA (red, closed rhombuses, solid line), compared to the expected piecewise-linear (black, closed circles) behavior (the latter is obtained by linear interpolation of the eLSDA energies at integer electron values); approach (ii) -- with LSDA (blue, open squares, dashed line) and eLSDA (red, open rhombs, dashed line), compared to the expected piecewise-linear (black, open circles, dashed line) behavior (the latter is obtained by linear interpolation of the eLSDA energies at integer electron values).}
  	\label{fig.E_N__F__2app} 
\end{figure}

In Fig.~\ref{fig.E_N__F__2app} we compare the $E(N)$ curves of the F atom, obtained with LSDA and eLSDA, for both approaches (i) and (ii). For LSDA, we find that for $N>9.8$ an unbound solution is obtained with approach (ii). For eLSDA, approach (ii) yields a convex energy curve, which approaches the same energy value as the LSDA curve at $N=10$. For the other atoms, namely H, Li, and C, a similar picture is obtained.

\begin{figure}[H]
	\centering
	\includegraphics[scale=0.6]{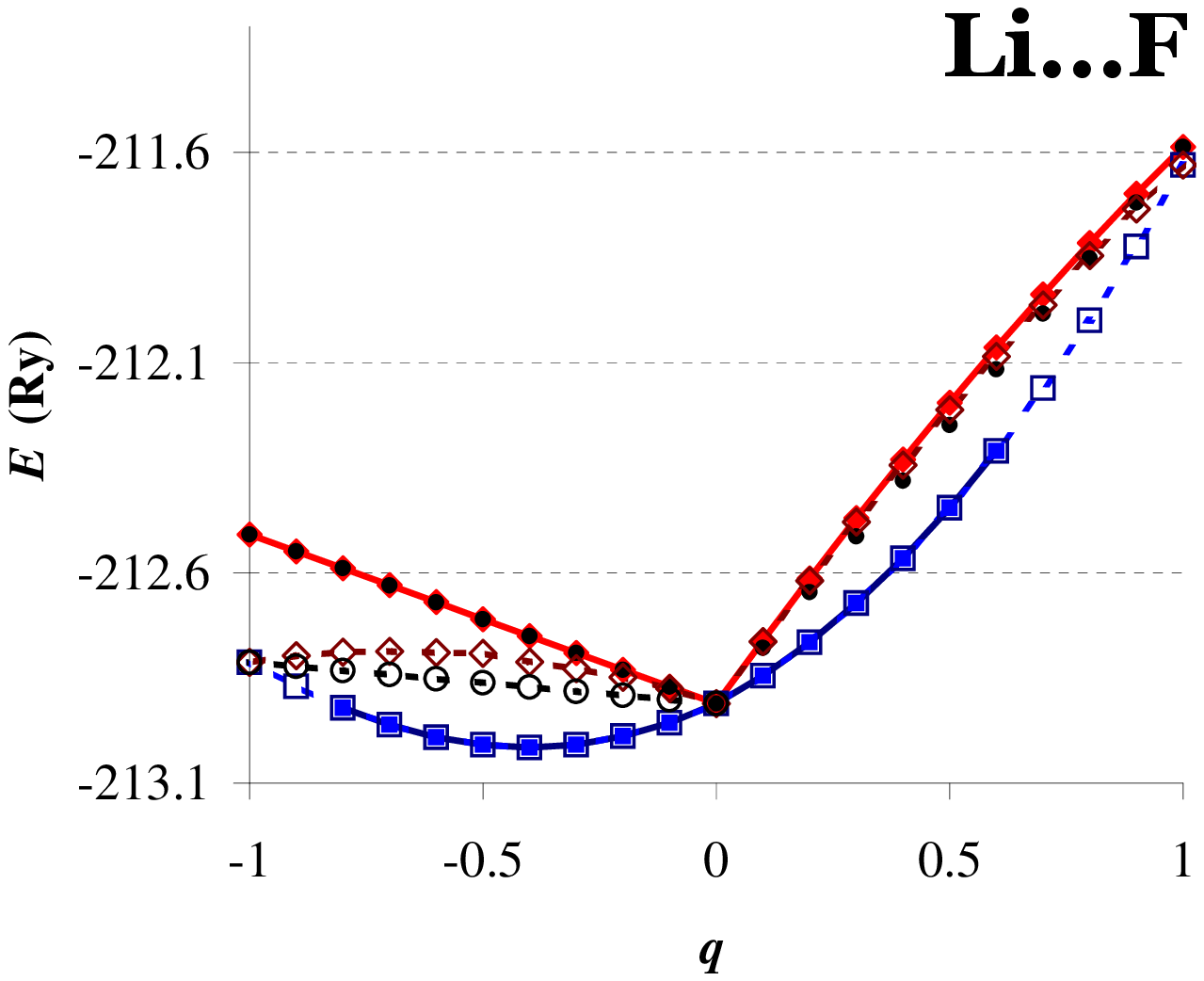}
   	\caption{Energy as a function of $q$ for the Li...F molecule, obtained using approaches (i) and (ii). All symbols are as in Fig.~\ref{fig.E_N__F__2app}. The piecewise-linear behavior in approach (ii) is plotted only for $q<0$, for clarity }
  	\label{fig.E_q_LiF__2app} 
\end{figure}

In Fig.~\ref{fig.E_q_LiF__2app} we present the $E(q)$ curves for the Li...F molecules, obtained with both approaches. We clearly see that the spurious LSDA energy minimum at $q_0 = -0.4$ is eliminated by the eLSDA  calculations performed in both approaches (i) and (ii). Instead, they produce a non-analytical minimum at $q=0$, as required.

We note that employment of approach (ii) in the calculations included several significant numerical difficulties. First, for the H and Li negative ions we found that if using \darsec 's standard initial guess for the density, for a certain $N$ the total energy jumps abruptly upwards and the energy levels drop downwards, yielding an artificial solution that is bound on one hand, but is higher in energy than that for the neutral atom. For Li, this happened at $N_1=3.346$. We relate this phenomenon to the fact that the numerical grid used in the calculation is not converged. To overcome the appearance of this solution, we had to restart our calculations with an initial guess taken from a calculation with $N<N_1$. 

Second, for the C and F negative ions, the self-consistent cycle could not converge when using the KLI approximation, due to the proximity of the $p$-levels in these atoms and frquent changes in their order during the self-consistent cycle. To overcome this difficulty, we had to use the S-iteration method~\cite{KumPer03a,KumPer03b}. Since the latter method is numerically sensitive, it is usually employed only after performing an initial step, e.g., of KLI. In our case, we used the simple Slater potential~\cite{Primer,KueKronik08} for the first iterations.

To summarize, a principal deficiency of the LSDA and eLSDA, i.e., not binding negative ions, raises numerous technical difficulties that obscures the discussion of the fractional dissociation problem. Ways to overcome them have been discussed above. Note that while relevant for neutral well-separated molecules, calculations of negative ions are not required for positively charged molecules. The latter have been considered in our work especially for illuminating the fractional dissociation problem in systems where binding problems are absent. As shown above, in positively charged molecule the fractional dissociation problem is removed by the ensemble treatment just as well as in neutral systems.

\bibliography{bibliography}